\documentclass[11pt]{article}              

\usepackage[margin=20mm]{geometry}
\usepackage{graphicx}
\usepackage{colortbl}
\usepackage[FIGBOTCAP]{subfigure}
 
\usepackage{fancyhdr}
\usepackage{amssymb,amsfonts,amsmath,amsthm}
\usepackage{natbib}

\usepackage{authblk}
	
\graphicspath{{./}}

\usepackage{bbm}
\usepackage{booktabs} 
\usepackage{multirow} 

\usepackage{color}

\usepackage{lineno}

\newcommand{\Rr}{\mathbb{R}}
\newcommand{\Nn}{\mathbb{N}}
\newcommand{\cD}{\mathcal{D}}
\newcommand{\ie}{\emph{i.e. }}
\newcommand{\eg}{\emph{e.g. }}
\newcommand{\prt}[1]{\left(#1\right)}		
\newcommand{\acc}[1]{\left\{#1\right\}}		
\newcommand{\bra}[1]{\left[ #1 \right]}

\newcommand{\var}[1]{\text{Var}\bra{#1}}
\newcommand{\uF}{\overline{F}}
\newcommand{\lF}{\underline{F}}
\newcommand{\ve}[1]{\boldsymbol{#1}}
\newcommand{\vt}{\ve{\theta}}
\newcommand{\vx}{\ve{x}}
\newcommand{\vX}{\ve{X}}
\newcommand{\cm}{\mathcal{M}}
\newcommand{\vC}{\ve{C}}

\newcommand{\cA}{\mathcal{A}}

\newcommand{\ua}{\ve{\alpha}}
\newcommand{\cY}{\mathcal{Y}}

\newcommand{\vcc}{\ve{\mathfrak{c}}}

\newcommand{\vT}{\ve{\Theta}}

\newcommand{\vV}{\ve{V}}
\newcommand{\vv}{\ve{v}}

\newcommand{\vchi}{\ve{\chi}}

\newcommand{\tr}{^{\textsf T}}

\newcommand{\ub}{\ve{\beta}}
\newcommand{\vtau}{\ve{\tau}}

\newcommand{\vChi}{\ve{\mathcal{X}}}
\newcommand{\vF}{{\bf F}}

\newcommand{\cI}{\mathcal{I}}
\newcommand{\cU}{\mathcal{U}}
\newcommand{\vVf}{\ve{\mathfrak{V}}}
\newcommand{\vvf}{\ve{\mathfrak{v}}}
\newcommand{\Wf}{\mathfrak{W}}
\newcommand{\nph}{n_{ph}}
\newcommand{\Ii}{\mathbb{I}}
\newcommand{\eqdef}{\stackrel{\text{def}}{=}}

\graphicspath{{./},{./Figures/}}

\begin{document}
\title{Global sensitivity analysis in the context of imprecise probabilities (p-boxes) using sparse polynomial chaos expansions} 

\author[1]{R. Sch\"obi}\author[1]{B. Sudret} 

\affil[1]{Chair of Risk, Safety and Uncertainty Quantification,
    ETH Zurich, Stefano-Franscini-Platz 5, 8093 Zurich, Switzerland}
%

\date{}
\maketitle

\abstract{Global sensitivity analysis aims at determining which
  uncertain input parameters of a computational model primarily drives
  the variance of the output quantities of interest. Sobol' indices
  are now routinely applied in this context when the input parameters
  are modelled by classical probability theory using random variables.
  In many practical applications however, input parameters are affected
  by both aleatory and epistemic (so-called polymorphic) uncertainty,
  for which imprecise probability representations have become popular in
  the last decade. In this paper, we consider that the uncertain input
  parameters are modelled by parametric probability boxes (p-boxes). We
  propose interval-valued (so-called imprecise) Sobol' indices as an
  extension of their classical definition. An original algorithm based
  on the concepts of augmented space, isoprobabilistic transforms and
  sparse polynomial chaos expansions is devised to allow for the
  computation of these imprecise Sobol' indices at extremely low cost.
  In particular, phantoms points are introduced to build an experimental
  design in the augmented space (necessary for the calibration of the
  sparse PCE) which leads to a smart reuse of runs of the original
  computational model. The approach is illustrated on three analytical
  and engineering examples which allows one to validate the proposed
  algorithms against brute-force double-loop Monte Carlo simulation.
  \\[1em]

  {\bf Keywords}: uncertainty quantification -- - global sensitivity
  analysis -- - probability-boxes -- - imprecise Sobol' indices --
  sparse polynomial chaos expansions}

\maketitle


\section{Introduction}

Computational simulation tools, such as finite element models (FEM), are the most popular approach to model complex systems and processes in modern engineering. Such simulation tools map a set of input parameters describing the system and its environmental and operational conditions through a computational model to a so-called quantity of interest (QoI), \eg performance indicators. The input parameters are often not perfectly known, due to noisy measurements, expert judgement, or intrinsic variability. Hence, each variable is modelled by an uncertainty model, as an example by a probability distribution. The uncertainty propagates through the computational model and results in an uncertain QoI.

Sensitivity analysis (SA) examines the impact of the uncertainty in the input vector of a computational model onto the uncertainty in the QoI. This is of importance in practice where the relation between the input variables and the QoI is implicitly defined by a complex computer code. A large number of SA methods can be found in the literature, as reported in the reviews in \cite{Saltelli2008,Helton2006,XuGertner2008}. The methods can be separated broadly into two classes, namely \emph{local} and \emph{global} SA methods. Local SA examines how small variations in the input variables affect the QoI, whereas global SA focuses on the entire variation of the input variables. In this paper, we look at global SA methods. 

Traditionally, global SA methods are developed in the context of probability theory, \ie the uncertainty in the input variables is modelled by probability distributions. However, a common situation in practice is to have only scarce or incomplete knowledge to characterize the uncertainty. This leads to the concept of epistemic uncertainty (representing the lack of data and lack of knowledge) alongside with aleatory uncertainty (natural variability) as a source of uncertainty. For such cases, \emph{imprecise probability} frameworks have been developed as a generalization of probability theory. Among these frameworks are Bayesian hierarchical models \cite{Gelman2009}, probability-boxes \cite{Ferson1996}, and fuzzy distributions \cite{Moller2004}. 

The number of publications for \emph{imprecise sensitivity analysis} (ISA) remains small. Li \& Mahadevan \cite{Li2016a,Li2016b}, Sankararaman \& Mahadevan \cite{Sankararaman2013a}, and Krzykacz-Hausmann \cite{Krzykacz-Hausmann2006} describe global SA methods in the presence of Bayesian hierarchical models. Pinching and imprecise Sobol' indices are discussed in \cite{Oberguggenberger2005,Oberguggenberger2009}. Pinching of probability-boxes is also elaborated in \cite{Ferson2006}. Helton et al. \cite{Helton2006a} discuss variance-based algorithm in the context of evidence theory.

SA methods, and in particular ISA methods, usually require a large number of evaluations of the computational model for different realizations of the input vector. Hence, such analyses become intractable when the computational model is an expensive-to-evaluate function. \emph{Meta-models}, a.k.a. \emph{surrogate models} and \emph{emulators}, are a popular solution to lower the computational costs by approximating the computational model by a cheap-to-evaluate analytical function. Examples include polynomial chaos expansions (PCE) \citep{Xiu2002,GhanemBook2003}, Kriging (a.k.a. Gaussian process models) \citep{Sacks1989,Santner2003}, support vector machines \citep{Gunn1998,Smola2006}, and artificial neural networks in the context of reliability analysis \citep{Hurtado2001,Schueremans2005}. 

A naive use of meta-models consists in the implementation of a sequential algorithm:  first, calibrate a meta-model and subsequently conduct SA (or ISA) on the approximation. There are, however, more elaborate ways to use meta-models in the context of SA. As an example, based on polynomial chaos expansions (PCE) \cite{GhanemBook2003}, Sudret \cite{Sudret2006d,Sudret2008c} computes Sobol' indices and Sudret \& Mai \cite{Sudret2015a} compute derivative-based global sensitivity measures (DGSM) analytically, see also Alexanderian et al \cite{Alexanderian2012a} and Le~Gratiet et al \cite{SudretHandbookUQ}. Konakli \& Sudret \cite{KonakliRESS2016} also obtain analytical expressions for Sobol' indices when the surrogate model is a low-rank tensor representation. However, the meta-model-based approaches are to-date limited to the probabilistic context. 

In this paper, the standard Sobol' indices \cite{Sobol1993} are first extended to the context of probability-boxes in order to account for mixed epistemic and aleatory uncertainty in global sensitivity analysis. Moreover, making use of the PCE-based Sobol' indices \cite{Sudret2008c} as well as a recently published variation of PCE \cite{SchoebiICASP2015}, a novel approach is proposed to compute imprecise Sobol' indices efficiently. Applying the proposed approach, ISA are expected to become tractable for realistic engineering problems. 

This paper is organized as follows. Section~\ref{sec:rem} recapitulates the basics of PCE and Sobol' indices in the standard setting of pure probabilistic modelling of input variables. Section~\ref{sec:imp} defines probability-boxes and derives the definition of imprecise Sobol' indices. In Section~\ref{sec:sma}, a smart PCE model is introduced, which is later used to compute the imprecise Sobol' indices in Section~\ref{sec:isp}. Three applications illustrate the workings of the proposed approach in Section~\ref{sec:appl}. The paper is terminated with conclusions in Section~\ref{sec:conc}.

\section{Polynomial chaos expansions for global sensitivity analysis} \label{sec:rem}
\subsection{Computational model}
%
Let us define a computational model $\cm$ as a mapping from the $M$-dimensional input space $\cD_{\vX}$ to the one-dimensional output space $\cD_Y$:
\begin{equation} \label{eq:cm}
\cm:\ \vx\in\cD_{\vX}\subset \Rr^M \mapsto y\in\cD_Y\subset\Rr.
\end{equation}
Herein, we assume that the model predicts a \emph{scalar} quantity of interest $y$ once it is fed with a vector of input parameters $\vx$. Extensions to vector-valued models are straightforward by applying the proposed methodology to each component separately. 

Due to uncertainties in the input vector, the latter is represented by a random vector $\vX$ with joint cumulative distribution function (CDF) $F_{\vX}$. The components of $\vX=\prt{X_1,\ldots,X_M}\tr$ are assumed independent for the sake of simplicity throughout this paper. The model response becomes a random variable $Y$ obtained by propagating the uncertainties in the input random vector $\vX$ through the computational model $\cm$. 

Uncertainty propagation is typically conducted by sampling-based algorithms such as Monte Carlo simulation \citep{Robert2004}. When the computational model is an expensive-to-evaluate function, the estimation of $Y$ is costly due to the repeated evaluations of the computational model. Hence, a number of techniques have been proposed in the literature for surrogating $\cm$. In the following section, Polynomial Chaos Expansions \citep{Xiu2002,GhanemBook2003,Soize2004} are introduced. 

\subsection{Sparse Polynomial Chaos Expansions} \label{sec:pce}
\subsubsection{Polynomial approximation}
Polynomial Chaos Expansions (PCE) approximate the computational model $\cm$ by a series of multivariate polynomials that are orthogonal to the distributions of the input variables \citep{GhanemBook2003,Soize2004}:
\begin{equation} \label{eq:cmpce}
Y\approx \cm^{(\text{P})}\prt{\vX} = \sum_{\ua\in\cA} a_{\ua} \psi_{\ua}\prt{\vX},
\end{equation}
where $\psi_{\ua}\prt{\vX}$ are multivariate orthonormal polynomials indexed by multi-indices $\ua=\prt{\alpha_1,\ldots,\alpha_M}$, ${\cA\subset\Nn^M}$ is a finite set of such indices (so-called truncation scheme), and $a_{\ua}\in\Rr$ are expansion coefficients. To derive the orthonormal polynomial basis, one proceeds as follows. Since the components of $\vX$ are assumed independent, its joint probability density function (PDF) is the product of its marginals. Then, a functional inner product for each marginal PDF $f_{X_i}$ is defined by:
\begin{equation}
\langle \phi_1,\phi_2\rangle_i = \int_{\cD_{X_i}} \phi_1(x)\, \phi_2(x)\ f_{X_i}(x)\ \text{d}x,
\end{equation}
for any two functions $\acc{\phi_1,\phi_2}$ such that the integral exists. For each variable $X_i$, an orthonormal polynomial basis can be constructed that satisfies \citep{Xiu2002}:
\begin{equation}
\langle P_j^{(i)}, P_k^{(i)} \rangle = \int_{\cD_{X_i}} P_j^{(i)}(x)\, P_k^{(i)}(x)\, f_{X_i}(x)\, \text{d}x = \delta_{jk},
\end{equation}
where $\acc{P_j^{(i)}, P_k^{(i)}}$ are two univariate polynomials in the $i^{\text{th}}$ variable and $\delta_{jk}$ is the Kronecker symbol which is equal to $1$ for $j=k$ and equal to $0$ otherwise. Xiu \& Karniadakis \citep{Xiu2002} summarize orthogonal bases for classical probability distributions.  

The multivariate polynomials are then obtained by tensor product of the univariate ones:
\begin{equation}
\psi_{\ua}\prt{\vx} = \prod_{i=1}^{M} P_{\alpha_i}^{(i)}\prt{x_i}. 
\end{equation}
Through the tensor product construction, the orthogonality property is kept, meaning that:
\begin{equation}
\mathbb{E}\bra{\psi_{\ua}\prt{\vX}\, \psi_{\ub}\prt{\vX}} \eqdef \int_{\cD_{\vx}} \psi_{\ua}\prt{\vx}\, \psi_{\ub}\prt{\vX}\, f_{\vX}\prt{\vx}\, \text{d}\vx = \delta_{\ua\ub},
\end{equation}
where $\delta_{\ua\ub}$ is the Kronecker delta with $\delta_{\ua\ub}=1$ for $\ua=\ub$ and $\delta_{\ua\ub}=0$ otherwise. 

\subsubsection{Truncation schemes}
The efficiency of PCE models greatly depends on the choice of the set of multi-indices $\cA$ in Eq.~(\ref{eq:cmpce}). A popular strategy to choose $\cA$ consists in upper-bounding the total degree of polynomials $\sum_{i=1}^{M} \alpha_i$ to a maximum value $p$ or in using hyperbolic index sets \citep{BlatmanPEM2010}:
\begin{equation} \label{eq:hyper}
\cA_q^{M,p} = \acc{\ua\in\Nn^M:\ ||\ua||_q\leq p}, \qquad ||\ua||_q = \prt{\sum_{i=1}^{M} \alpha_i^q}^{1/q},
\end{equation}
where $0<q\leq 1$ is a parameter and $p$ is the maximal total degree of the univariate polynomials retained. 

\subsubsection{Computation of coefficients $a_{\ua}$}
After defining the set of candidate polynomials, the coefficients $a_{\ua}$ in Eq.~(\ref{eq:cmpce}) are to be determined. Based on a finite set of realizations of the input vector $\vX$, denoted by $\vChi=\acc{\vchi^{(1)},\ldots, \vchi^{(N)}}$ and called experimental design, and the corresponding response values $\cY = \acc{\cY^{(1)},\ldots, \cY^{(N)}}\eqdef \acc{\cm\prt{\vchi^{(1)}},\ldots, \cm\prt{\vchi^{(N)}}}$, the coefficients are calculated by discretized least-squares minimization \cite{Berveiller2006,SudretHDR}:
\begin{equation}
\widehat{\ve{a}} = \arg\min_{\ve{a}\in\Rr^{|\cA|}} \frac{1}{N}\sum_{i=1}^{N} \prt{\cY^{(i)} - \sum_{\ua\in\cA} a_{\ua}\, \psi_{\ua}\prt{\vchi^{(i)}}  }^2,
\end{equation}
where $n_{\cA} = |\cA|$ is the number of polynomials retained in the truncated expansion. The optimal coefficients $\widehat{\ve{a}}$ can be obtained by solving the linear system:
\begin{equation}
\widehat{\ve{a}} = \prt{\vF\tr\vF}^{-1}\vF\tr\cY.
\end{equation}
In this expression, the information matrix $\vF$ contains the values of all selected polynomials at all points of the experimental design, namely: 
\begin{equation}
F_{ij}=\psi_j\prt{\vchi^{(i)}}, \qquad i=1,\ldots,N, \qquad j=1,\dots,n_{\cA}.
\end{equation}

\subsubsection{Polynomial selection}
Typically for smooth functions, a small number of polynomials is sufficient to accurately represent the output of the computational model. Thus, a further reduction of the set of predictors is possible. A number of methods are available in the literature, including the \emph{least absolute shrinkage and selection operator} (LASSO) \citep{tibshirani1996}, \emph{least angle regression} (LAR) \citep{Efron2004}, \emph{orthogonal matching pursuit} (OMP) \citep{Pati1993,Mallat1993}, and \emph{Bayesian compressive sensing} \citep{Sargsyan2014}. Based on LAR, Blatman \& Sudret \cite{BlatmanJCP2011} proposed the LARS algorithm which allows for an efficient selection of a small number of polynomials out of the candidate set $\cA_q^{M,p}$ (see also Eq.~(\ref{eq:hyper})). For a thorough discussion of the LARS algorithm, the reader is referred to \cite{BlatmanJCP2011,BlatmanThesis}. 

\subsection{Sobol' indices}

The so-called \emph{Sobol' decomposition} represents the computational model by a series of summands of increasing dimension \citep{Sobol1993}:
\begin{multline} \label{eq:decomp}
\cm(\vx) = \cm_0 + \sum_{i=1}^{M} \cm_i(x_i) + \sum_{1\leq i<j\leq M} \cm_{ij}(x_i,x_j) + \ldots\\
+ \sum_{1\leq i_1<\ldots<i_s\leq M} \cm_{i_1\ldots i_s}\prt{x_{i_1},\ldots,x_{i_s}} + \ldots + \cm_{1,2,\ldots,M}(\vx),
\end{multline}
where $\cm_0$ is a constant (mean value of the function) and where it is imposed that the integral of each summand $\cm_{i_1,\ldots,i_s}\prt{x_{i_1},\ldots,x_{i_s}}$ over any of its arguments is zero:
\begin{equation} \label{eq:int0}
\int_{\cD_{X_{k}}} \cm_{i_1\ldots i_s}(x_{i_1},\ldots,x_{i_s})\, f_{X_k}\prt{x_k}\,\text{d}x_{k} = 0,\qquad 1\leq i_1<\ldots < i_s\leq M, \qquad k\in\acc{i_1,\ldots,i_s}.
\end{equation}
The mean value of the function can be computed as:
\begin{equation}
\cm_0 = \int_{\cD_{\vX}} \cm(\vx)\, f_{\vX}\prt{\vx} \, \text{d}\vx.
\end{equation}
Due to the constraints imposed in Eq.~(\ref{eq:int0}), the summands are orthogonal to each other so that \citep{Homma1996}:
\begin{equation} \label{eq:ortho}
\int_{\cD_{\vX}} \cm_{i_1\ldots i_s}(x_{i_1},\ldots,x_{i_s}) \, \cm_{j_1\ldots j_t}(x_{j_1,\ldots,j_t})\, f_{\vX}\prt{\vx} \, \text{d}\vx = 0,\qquad\acc{i_1,\ldots,i_s} \neq \acc{j_1,\ldots, j_t}.
\end{equation}

Considering now that the input parameters are modelled by independent random variables, the \emph{total variance} of the computational model is defined as:
\begin{equation}
D = \var{\cm(\vX)} = \int_{\cD_{\vX}} \cm^2(\vx) \, f_{\vX}\prt{\vx}\,\text{d}\vx - \cm_0^2.
\end{equation}
By using the properties of Eq.~(\ref{eq:ortho}) in Eq.~(\ref{eq:decomp}), one obtains the following decomposition of the variance:
\begin{equation} \label{eq:dd}
D = \sum_{i=1}^{M} D_i + \sum_{1\leq i< j \leq M} D_{ij} + \ldots + D_{1,2,\ldots,M},
\end{equation}
where the \emph{partial variances} are computed as:
\begin{equation} 
D_{i_1,\ldots,i_s} = \int_{\cD_{\vX}} \cm_{i_1\ldots i_s}^2(x_{i_1},\ldots,x_{i_s}) \, f_{\vX}\prt{\vx}\, \text{d}\vx, \qquad 1\leq i_1 < \ldots < i_s \leq M, \qquad s=1,\ldots,M.
\end{equation}

Making use of this decomposition, the Sobol' indices are defined as the relative partial variances:
\begin{equation} \label{eq:sobol}
S_{i_1\ldots i_s} = \frac{D_{i_1\ldots i_s}}{D} = \frac{D_{i_1\ldots i_s}}{\sum_{i=1}^{M} D_i + \sum_{1\leq i< j \leq M} D_{ij} + \ldots + D_{1,2,\ldots,M}},
\end{equation}
and by virtue of Eq.~(\ref{eq:dd}), it holds that:
\begin{equation}
\sum_{i=1}^{M} S_i + \sum_{1\leq i < j \leq M} S_{ij} + \ldots + S_{1,2,\ldots,M} = 1.
\end{equation}
The Sobol' indices measure the amount of the total variance coming from the uncertainties in a set of input parameters. In practice, it is common to compute the \emph{first order} indices $S_i$, which measure the influence of each parameter taken separately. Higher order indices account for the interactive contributions to the total variance. In this sense, the \emph{total sensitivity indices} $S_i^{(T)}$ are defined as:
\begin{equation} \label{eq:sobolT}
S_{i}^{(T)} = \sum_{\acc{ i_1,\ldots,i_s } \supset \acc{i}} \frac{D_{i_1\ldots i_s}}{D}.
\end{equation}
In other words, $S_{i}^{(T)} = 1 - S_{-i}$, where $S_{-i}$ is the sum of all $S_{i_1\ldots i_s}$ that do not include index $i$.

\subsection{PCE-based Sobol' indices}

Let us consider now that an appropriate approximation of the computational model $\cm$ exists in the form of a truncated PCE, as seen in Eq.~(\ref{eq:cmpce}). Then, a set of multi-indices $\mathcal{I}_{i_1,\ldots,i_s}$ can be defined so that only the indices $i_1,\ldots,i_s$ are non-zero, \emph{i.e.}:
\begin{equation}
\cI_{i_i,\ldots,i_s} = \acc{ \ua\in\cA: \begin{array}{l}
\alpha_k > 0 \quad \text{for all} \quad k\in \acc{ i_1,\ldots,i_s } \\ 
\alpha_k = 0 \quad \text{for all} \quad k\not\in \acc{ i_1,\ldots, i_s }
\end{array}  }.
\end{equation}
The elements of Eq.~(\ref{eq:cmpce}) can be reordered according to the decomposition in Eq.~(\ref{eq:decomp}) \citep{Sudret2008c}:
\begin{multline} \label{eq:pceso}
\cm^{(\text{P})}(\vx) = a_0 + \sum_{i=1}^{N} \prt{ \sum_{\ua \in \cI_i} a_{\ua} \psi_{\ua}(x_i) } 
 + \sum_{1\leq i_1 < i_2 \leq M} \prt{ \sum_{\ua\in\cI_{i_1,i_2}} a_{\ua} \psi_{\ua}(x_{i_1},x_{i_2}) } + \ldots \\
+ \sum_{1\leq i_1<\ldots<i_s \leq M} \prt{ \sum_{\ua\in\cI_{i_1,\ldots,i_s}} a_{\ua} \psi_{\ua}(x_{i_1},\ldots,x_{i_s}) }
+ \ldots + \sum_{\ua\in\cI_{1,2,\ldots,M}} a_{\ua}\psi_{\ua}(x_1,\ldots,x_M).
\end{multline}
Then, the summands of Eq.~(\ref{eq:pceso}) can be identified as summands of the Sobol' decomposition:
\begin{equation} \label{eq:cmi}
\cm_{i_1\ldots i_s}(x_{i_1},\ldots,x_{i_s}) = \sum_{\ua\in\cI_{i_1,\ldots,i_s}} a_{\ua} \psi_{\ua}(x_{i_1},\ldots,x_{i_s}).
\end{equation}
Due to the property of uniqueness, it can be concluded that Eq.~(\ref{eq:pceso}) is indeed the Sobol' decomposition of the PCE model. Furthermore, due to the orthonormality of the PC basis, the partial variance $D_{i_1\ldots i_s}$ is readily obtained from Eq.~(\ref{eq:cmi}) as:
\begin{equation}
D_{i_1\ldots i_s} = \sum_{\ua\in\cI_{i_1,\ldots,i_s}} a_{\ua}^2.
\end{equation}
The corresponding \emph{polynomial-chaos-based Sobol' indices}, denoted by $S^{(\text{P})}_{i_1\ldots i_s}$, are then defined, as originally proposed in \cite{Sudret2006d}, as:
\begin{equation}
S^{(\text{P})}_{i_1\ldots i_s} =  \left. \sum_{\ua \in\cI_{i_1,\ldots,i_s}} a^2_{\ua}  \middle/ \sum_{\ua \in \cA,\ \ua \neq \ve{0}} a^2_{\ua} \right..
\end{equation}
The \emph{total PC-based sensitivity indices} are then computed analogously to Eq.~(\ref{eq:sobolT}):
\begin{equation} \label{eq:sobolt}
S^{(T)(\text{P})}_{i} = \sum_{\acc{ i_1,\ldots,i_s } \supset \acc{i}  } S^{(\text{P})}_{i_1,\ldots,i_s}. 
\end{equation}

As seen in this section, the computation of the Sobol' indices for a PCE model boils down to working with the coefficients of the PC-expansion, which is a simple operation at almost no additional computational cost. Hence, PCE-based Sobol' indices are obtained by simple post-processing of an existing PCE model. 

\section{Imprecise Sobol' indices} \label{sec:imp}
\subsection{Free and parametric p-boxes}
%
Sobol' indices can be computed by the procedure described in the above section as soon as the input parameters are modelled by well-defined probability density functions. In practice however, probability theory is often not fully appropriate to characterize the uncertainty in the system. Probability-boxes (p-boxes) pose an intuitive generalization of probability theory. Instead of representing $X$ by a single CDF, lower and upper bounds denoted by $\lF_X$ and $\uF_X$ are introduced \citep{Ferson1996,Ferson2004}. For any value of $x\in\cD_X$, the true-but-unknown CDF lies within these bounds such that $\lF_X(x)\leq F_X(x) \leq \uF_X(x)$, which form a so-called p-box.

In the literature, two types of p-boxes can be found, namely \emph{free} and \emph{parametric} p-boxes. Free p-boxes are defined directly though the formulation of the bounds $\acc{\lF_X,\uF_X}$. Parametric p-boxes (a.k.a distributional p-boxes) are defined as cumulative distribution functions with interval-valued hyper-parameters:
$F_X(x) = F_X(x|\vt)$, where $\vt\in\cD_{\vT}\subset\Rr^{n_{\vt}}$,
and $\cD_{\vT}$ is the interval domain of the distribution parameters of dimension $n_{\vt} = |\vt|$. Precisely, $\cD_{\vT}= \bra{\underline{\theta}_1,\overline{\theta}_1}\times\ldots\times\bra{\underline{\theta}_{n_{\vt}},\overline{\theta}_{n_{\vt}}}$ denotes a hyper-rectangular domain, and $\underline{\theta}_k$ and $\overline{\theta}_k$ denote the lower and upper bounds of the interval $\bra{\underline{\theta}_k,\overline{\theta}_k}$ for the $k$-th parameter of the CDF of $X$. Boundary curves for parametric p-boxes can be defined as for free p-boxes as follows:
\begin{equation} \label{eq:pboxbounds}
\lF_X^{(p)}(x) = \min_{\vt\in\cD_{\vT}} F_X(x|\vt), \qquad \uF_X^{(p)}(x) = \max_{\vt\in\cD_{\vT}} F_X(x|\vt).
\end{equation}

\begin{figure}[!ht]
\centering
\subfigure[Free p-box \label{fig:pbox:free}]{
  \includegraphics[width=0.45\linewidth]{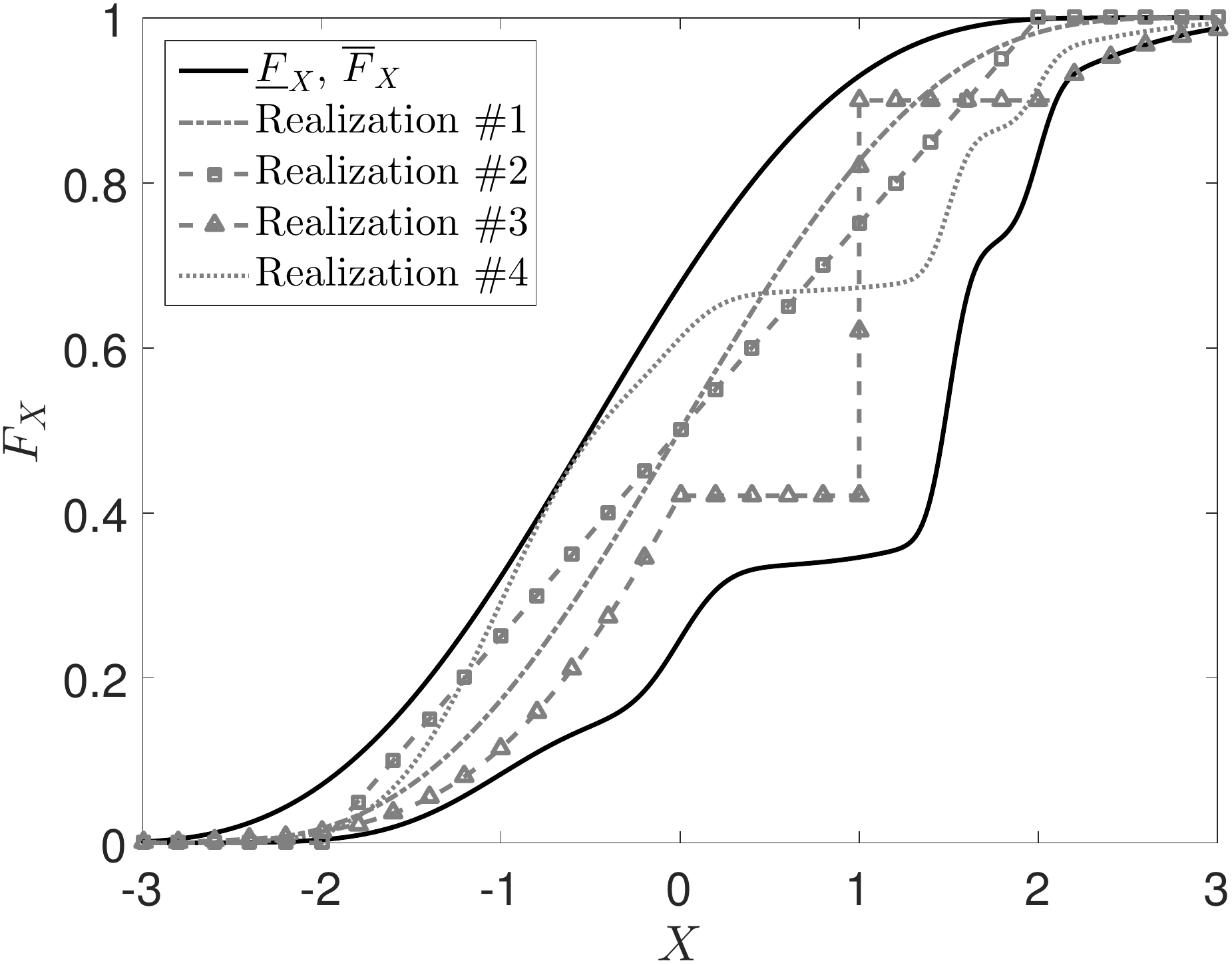}
}
\subfigure[Parametric p-box]{
  \includegraphics[width=0.45\linewidth]{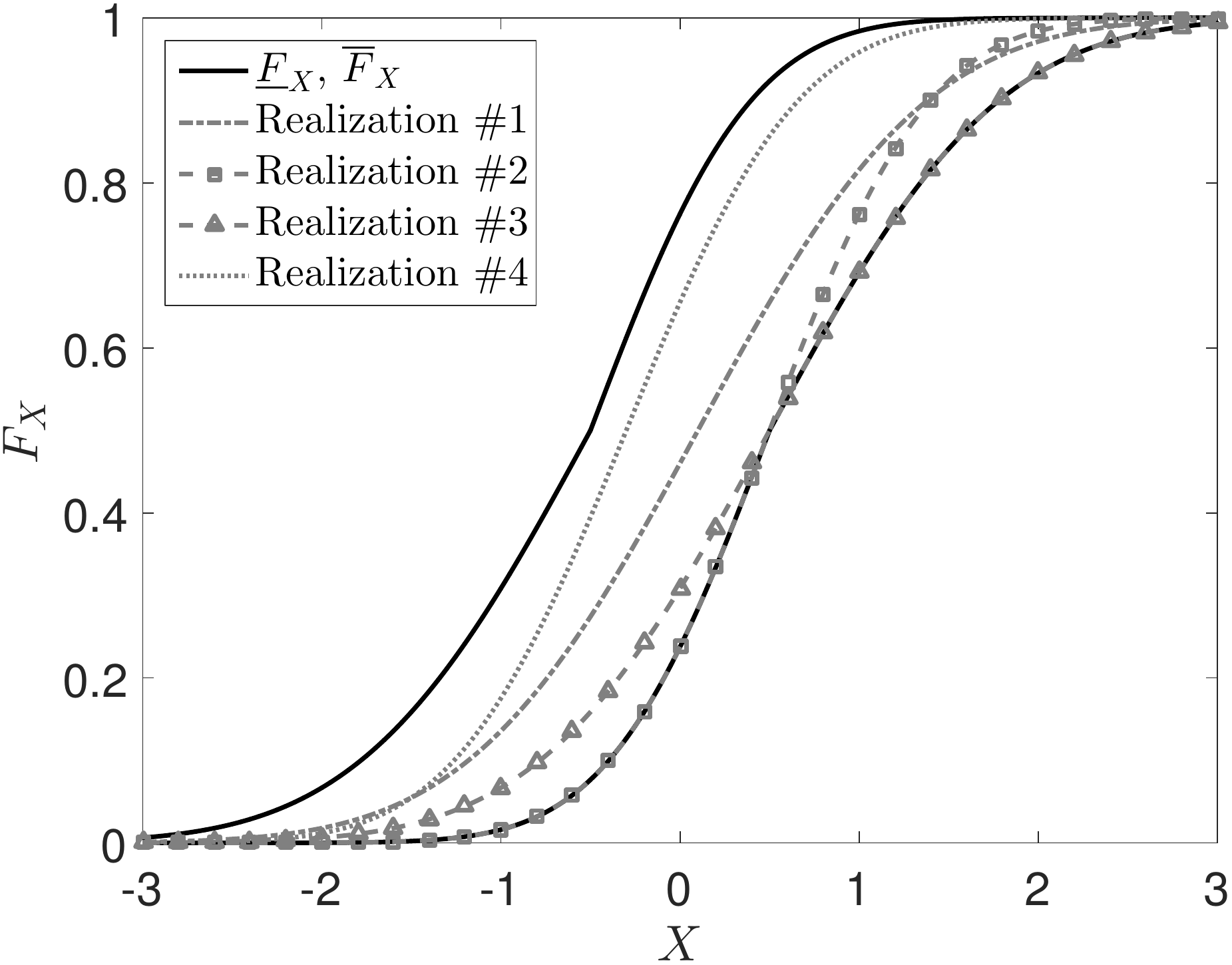}
}
\caption{\label{fig:pbox} Examples of the two types of p-boxes}
\end{figure}

Figure~\ref{fig:pbox} shows the intrinsic difference between free and parametric p-boxes. In both cases, only non-decreasing realizations of the true-but-unknown CDF are feasible. However, free p-boxes are a more general description of uncertainty than parametric p-boxes. This can be seen by the stair-case realizations, which are possible in free but not in parametric p-boxes (see also Realization \#3 in Figure~\ref{fig:pbox:free}).

In this paper, the focus lies on the parametric p-boxes due to their clear separation of aleatory and epistemic uncertainty: the distribution family represents the aleatory uncertainty, whereas the interval-valued hyper-parameters represent the epistemic uncertainty.  

\subsection{Imprecise Sobol' indices}
%
In the context of parametric p-boxes, uncertainty quantification is typically conducted by nested Monte Carlo simulations (nMCS) algorithms where an outer loop samples the parameters $\vt$ and an inner loop samples the conditional probability distribution $F_{\vX}\prt{\vx|\vt}$ \citep{Eldred2009,Chowdhary2013}. Taking this idea into the context of sensitivity analysis, the Sobol' indices can be computed for each realization $\vt^{(j)}$ and the corresponding $F_{\vX}\prt{\vx\middle|\vt^{(j)}}$. Each realization can result in a different set of Sobol' indices. Finally, the bounds of the imprecise Sobol' indices can be determined by analysing the set of realizations. 

In the sequel, we define imprecise Sobol' indices $S_{i_1\ldots i_s}$ as a pair of bounds obtained by minimizing (resp. maximizing) the classical Sobol' indices over the range of the hyper-parameters $\vt\in\cD_{\vT}$: 
\begin{equation} \label{eq:sobolminmax0}
\underline{S}_{i_1\ldots i_s} = \min_{\vt\in\cD_{\vT}} S_{i_1\ldots i_s}\prt{\vt}, \qquad \overline{S}_{i_1\ldots i_s} = \max_{\vt\in\cD_{\vT}} S_{i_1\ldots i_s}\prt{\vt},
\end{equation}
where $S_{i_1\ldots i_s} \in\bra{\underline{S}_{i_1\ldots i_s},\, \overline{S}_{i_1\ldots i_s}}$ are the interval-valued Sobol' indices and $S_{i_1\ldots i_s}\prt{\vt}$ is the corresponding Sobol' index for the conditional input distribution $F_{\vX}\prt{\vx|\vt}$. Similarly to Eq.~(\ref{eq:sobolt}), the bounds of the total Sobol' indices can be obtained by:
\begin{equation} \label{eq:sobolminmax}
\underline{S}_j^{(T)} = \min_{\vt\in\cD_{\vT}} S_j^{(T)}\prt{\vt}, \qquad \overline{S}_j^{(T)} = \max_{\vt\in\cD_{\vT}} S_j^{(T)}\prt{\vt}.
\end{equation} 

\subsection{PCE-based indices}
%
From the definitions above, it can be seen that computing the two bounds of imprecise Sobol' indices requires solving two optimization problems over the hyper-parameter space $\cD_{\vT}$. In order to speed up the process, we propose to use polynomial chaos expansions as presented in Section~\ref{sec:pce}. 

The use of the PCE meta-models would reduce the computational effort of computing Sobol' indices conditioned on a particular $\vt$ but does not remove the repeated computation of the variance decomposition for many $\vt\in\cD_{\vT}$ to solve the optimization problems. In other words, when adopting a brute force procedure, a large number of PCE models should be calibrated, each of them with its own experimental design, own evaluations of the costly computational model $\cm$, etc. Hence in the following, we introduce a smart PCE based on a \emph{single experimental design} that is sufficient to estimate any $S_i\prt{\vt}$ where $\vt\in\cD_{\vT}$. 

\section{Augmented polynomial chaos expansions for parametric distributions} \label{sec:sma}
\subsection{Augmented space}
%
Due to the clear separation of aleatory and epistemic uncertainty in the formulation of parametric p-boxes, these sources of uncertainty can be treated as separate entities, \ie $\vX$ and $\vT$. Hence, the QoI can be interpreted as a function of the augmented input vector $\prt{\vX,\vT}$. The corresponding map then reads:
\begin{equation}
W = \cm^{(\text{aug})}\prt{\vX,\vT},
\end{equation}
where the \emph{augmented computational model} is based on the original computational model as $\cm^{(\text{aug})}\prt{\vx,\vt}\equiv \cm\prt{\vx}$. 

\begin{figure}[!ht]
\centering
\subfigure[Original dependencies \label{fig:dep:a}]{
	\includegraphics[width=0.45\linewidth]{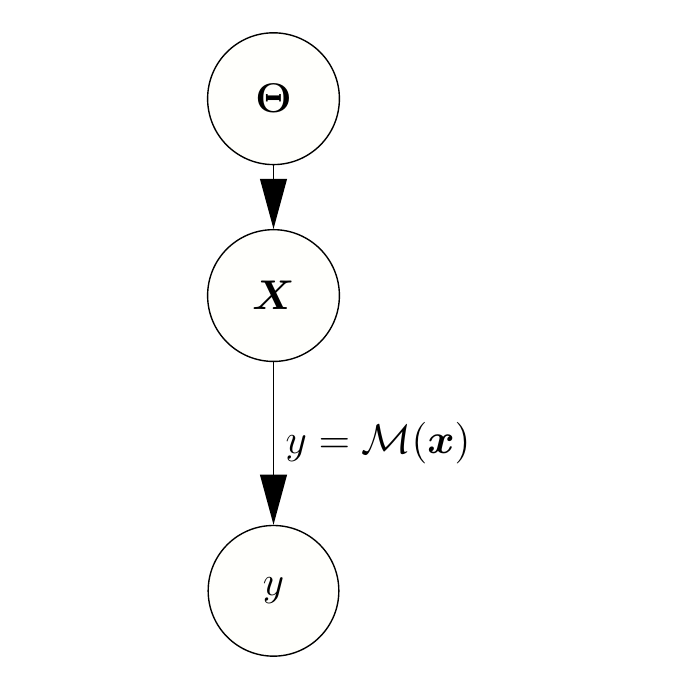}
}
\subfigure[Independent input variables \label{fig:dep:b}]{
	\includegraphics[width=0.45\linewidth]{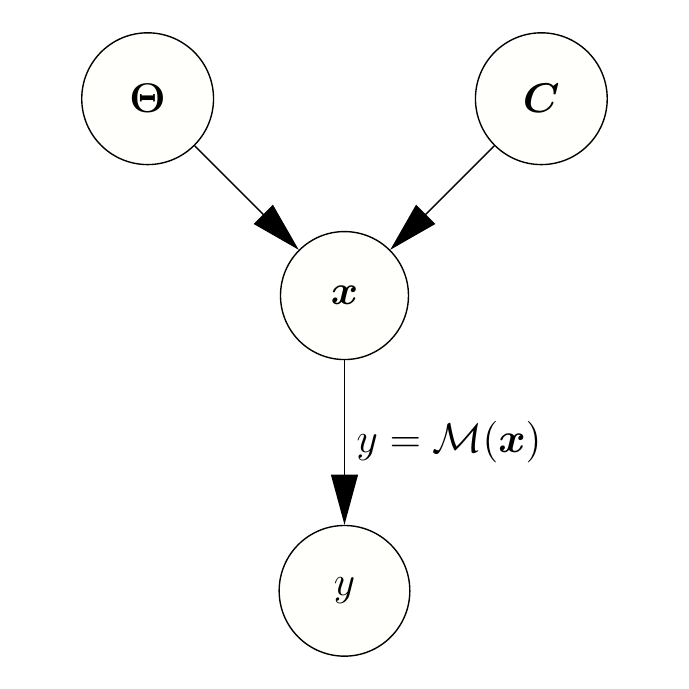}
}
\caption{Parametric p-box --  dependency structures before and after isoprobabilistic transform \label{fig:dep}}
\end{figure}

The components of the augmented input vector are dependent on each other because $\vX$ depends on $\vT$ as a result of the hierarchical formulation of the parametric p-box. The dependencies are visualized in Figure~\ref{fig:dep:a}. In this paper, however, the input variables to the computational model are assumed to be independent. In order to obtain independent variables in the augmented input vector, an isoprobabilistic transform is used that maps $\prt{\vX,\vT}$ to $\vV$, which shall be a vector with independent components. Assuming that the components $X_i$ are independent and that the parameter intervals in $\vT$ are independent too (\ie $\cD_{\vT}$ is a hyper-rectangular domain), the parametric p-box $\vX$ can be recast as:
\begin{equation} \label{eq:augmentedmodel:separate}
X_i = F_{X_i}^{-1}\prt{C_i|\vt_i}, \quad i=1,\ldots,M.
\end{equation} 
where $\vC=\prt{C_1,\ldots,C_M}$ is a vector of independent uniform distributions and where $C_i\sim\cU(0,1)$ describes the CDF value of $X_i$. Then, the augmented vector $\vV=\prt{\vC,\vT}$ is a vector with independent components that allows to define explicitly any realization of $\vX$ through Eq.~(\ref{eq:augmentedmodel:separate}).

The corresponding probabilistic graphical model transform is visualized in Figure~\ref{fig:dep:b}. Note that $\vx$ is used in lower case due to the deterministic transform of $\vV=\prt{\vC,\vT}\mapsto\vX$. In Figure~\ref{fig:dep:a}, the transform of $\vT$ to $\vX$ is not deterministic, hence, the upper case $\vX$.

\subsection{Augmented PCE}
%
The augmented computational model $\cm^{(\text{aug})}$ can be formulated as a function of vector $\vV$ as follows:
\begin{equation} \label{eq:ww}
W = \cm^{(\text{aug})}\prt{T\prt{\vV}},
\end{equation}
where $T$ denotes the isoprobabilistic transform from $\vV$ to $\prt{\vX,\vT}$. When interpreting the epistemic intervals in $\vT$ as uniform distributions, a sparse PCE model can be calibrated to meta-model the augmented computational model (analogously to Eq.~(\ref{eq:cmpce})): 
\begin{equation} \label{eq:augpce}
W\approx\cm^{(\text{PCE})}\prt{\vV} = \sum_{\ua\in\cA} a_{\ua} \ \psi_{\ua}\prt{\vV},
\end{equation}
where the dimensionality is $M = |\vV| = |\vC| + |\vT|\equiv |\vX| + |\vT|$. The meta-model should be trained with an experimental design $\vVf=\acc{ \vvf^{(1)},\ldots,\vvf^{(N)} }$  and the corresponding responses:
$$
\Wf  =\acc{\Wf^{(1)} = \cm^{(\text{aug})}\prt{T\prt{\vvf^{(1)}}},\ldots, \Wf^{(N)} = \cm^{(\text{aug})}\prt{T\prt{\vvf^{(N)}}}},
$$
by using Eq.~(\ref{eq:ww}). In order to efficiently calibrate a sparse PCE meta-model, the reader is referred to Section~\ref{sec:pce} for further details. Note that in this setup, the input vector $\vV$ is sampled in the same way for the $\vC$- and the $\vT$-components.

\subsection{Phantom points}
\subsubsection{Isoprobabilistic transform}
%
The number of dimensions in the augmented input space $\cD_{\vV}$ is larger than the one of the original input space $\cD_{\vX}$ due to the epistemic uncertainty. However, Eq.~(\ref{eq:augmentedmodel:separate}) leads to an interesting feature of the augmented PCE. For a given realization of $\vx\in\cD_{\vX}$, there is an infinite number of realizations $\vv$ that satisfy Eq.~(\ref{eq:augmentedmodel:separate}). In other words, the inverse operation of Eq.~(\ref{eq:augmentedmodel:separate}) is non-unique. This characteristics can be exploited when constructing the experimental design of the augmented PCE model. 

Consider again the experimental design $\vVf$. Each sample $\vvf^{(j)}$ can be transformed into an equivalent $\vchi^{(j_0)}$ using Eq.~(\ref{eq:augmentedmodel:separate}). A number of realizations $\vvf=\prt{\vcc,\vtau}$ can be generated by sampling $\vtau\in\cD_{\vT}$ and inserting them into the inverse isoprobabilistic transform to obtain realizations $\vcc$ that all correspond to the single point $\vchi^{(j_0)}$  in the original experimental design. In particular, for each dimension $i=1,\ldots,M$ and experimental design sample $j=1,\ldots,N$, $\nph$ samples can be generated from $\vt_i\in \cD_{\vT_i}$ and collected in $\acc{\vtau_i^{(j)(k)}, k=1,\ldots,\nph}$. Then, the inverse transform consists of evaluating simple CDFs and reads:
\begin{equation} \label{eq:nph}
\mathfrak{c}_i^{(j)(k)} = F_{X_i} \prt{ \chi_i^{(j)}\middle| \vtau_i^{(j)(k)}},
\end{equation}
where each experimental design sample point is $\vchi^{(j)} = \prt{\chi_1^{(j)},\ldots,\chi_M^{(j)}}$. The new components in dimension $i$ of vector $\vvf^{(j)}$ reads then:
\begin{equation}\label{eq:nphset}
\acc{\vvf_i^{(j)(k)} = \prt{  \mathfrak{c}_i^{(j)(k)},  \vtau_i^{(j)(k)}  },\  k=1,\ldots,\nph }.
\end{equation} 

When pursuing this procedure for all $i=1,\ldots,M$ dimensions of the input vector, new samples of the experimental design are generated. These samples are called \emph{phantom points} \citep{SchoebiICASP2015} because they contribute to the experimental design $\vVf$ without increasing the number of model evaluations $\cm$, which is once and for all $N$. Indeed, the $\nph$ points in Eq.~(\ref{eq:nphset}) all correspond to a \emph{single} $\vchi^{(j_0)}$ in the original experimental design, meaning a \emph{single run} $\cY^{(j_0)}=\cm\prt{\vchi^{(j_0)}}$. This has a large effect on the global efficiency when considering that model evaluations are dominating the total computational costs. The experimental design with phantom points and the corresponding responses then read:
\begin{equation}
\left\{
\begin{array}{l}
\vVf = \acc{\vvf^{(j)(k)},  \quad j=1,\ldots,N, \quad k=1,\ldots,\nph} \\ 
\\
\Wf = \acc{\Wf^{(j)(1)}=\ldots=\Wf^{(j)(\nph)}, \quad j=1,\ldots,N}
\end{array} 
\right\},
\end{equation}
which consists of $\nph\times N$ samples at a cost of exactly $N$ evaluations of the computational model $\cm$. 

\begin{figure}[!ht]
\centering
\includegraphics[width=0.45\linewidth]{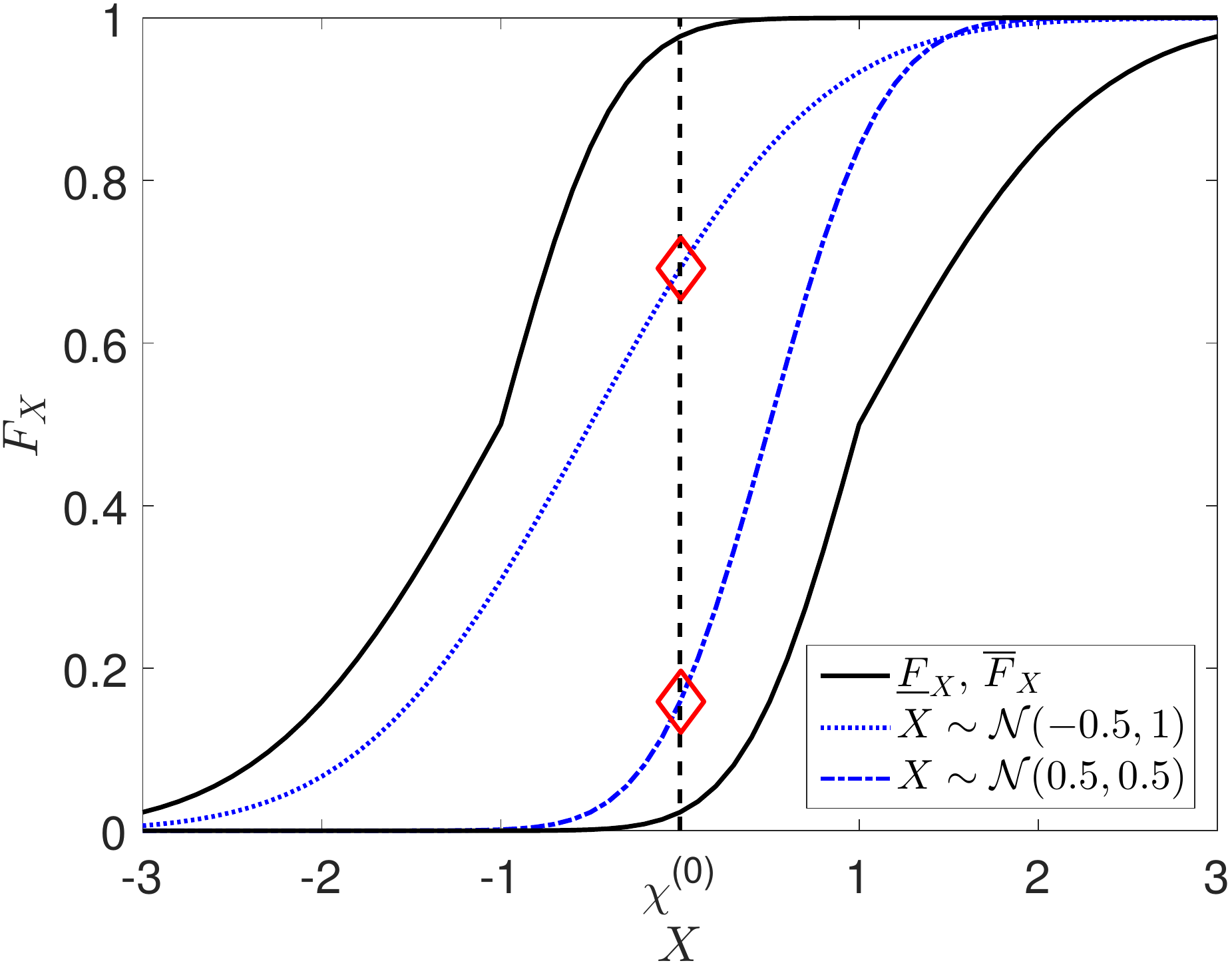}
\caption{\label{fig:phantom} Illustration of phantom points on a parametric p-box $X\sim\mathcal{N}\prt{\mu_X,\sigma_X}$, where $\mu_X\in\bra{-1,\,1}$ and $\sigma_X\in\bra{0.5,\,1.0}$.}
\end{figure}

For illustration purposes, consider a parametric p-box defined by a Gaussian distribution family with interval-valued mean and standard deviation. As an example, it is assumed that ${\mu_X\in\bra{-1,\,1}}$ and ${\sigma_X\in\bra{0.5,\,1.0}}$. Figure~\ref{fig:phantom} shows the bounds of the parametric p-box (see Eq.~(\ref{eq:pboxbounds})) as well as the experimental sample point $\chi^{(0)}=0$. Furthermore, the dotted and dashed-dotted blue lines show two realizations of the CDF in the parametric p-box, namely $\acc{\mu_X=-0.5,\, \sigma_X=1}$ and $\acc{\mu_X=0.5,\,\sigma_X=0.5}$, respectively. The phantom points corresponding to $\chi^{(0)}$ are then computed through the CDF of the Gaussian distribution:
\begin{eqnarray}
\vvf^{(0)(1)} &=& \acc{\mu_X = -0.5, \, \sigma_X = 1, \, c = F_{\mathcal{N}}\prt{\chi^{(0)}\middle| -0.5, 1}\approx 0.6915}, \\
\vvf^{(0)(2)} &=& \acc{\mu_X = 0.5, \, \sigma_X = 0.5, \, c = F_{\mathcal{N}}\prt{\chi^{(0)}\middle| 0.5, 0.5}\approx 0.1587},
\end{eqnarray}
which are highlighted by the red diamonds in Figure~\ref{fig:phantom}.

As a summary, a parametric p-box $F_x\prt{x\middle|\vt;\,\vt\in\cD_{\vT}}$ can be equivalently represented by $\prt{n_{\vt}+1}$ independent random variables in the augmented space. The extended model $\cm^{(\text{aug})}$ in the augmented space is formulated as a function of a $n_{\text{aug}}$-dimensional input vector, where $n_{\text{aug}} = M + \sum_{i=1}^{M} n_{\vt}^{(i)}$, $n_{\vt}^{(i)}$ being the number of parameters $\vt$ characterizing the $i^\text{th}$ p-box. However, using the ``phantom points'' trick, an experimental design of arbitrary size $\nph\times N$ can be computed in the augmented input space at the cost of only $N$ evaluations of the original model $\cm$. This leads to a remarkably efficient computational procedure. Further computational details are given below. 

\subsubsection{Case of unbounded distributions}
%
In the case of unbounded distributions, the phantom points can be generated as described above. However, the CDF used in Eq.~(\ref{eq:nph}) is a highly non-linear function due to its mapping of an unbounded to a bounded domain, \ie $X\in\bra{-\infty, \, \infty}\mapsto C\in\bra{0,\,1}$. In order to avoid the non-linearity in this mapping and hence expecting more accurate meta-models when using low-degree polynomials, an auxiliary variable $\xi$ is introduced replacing the uniform distribution $C$. Then, an auxiliary isoprobabilistic transform is formulated for the mapping $\prt{\xi,\vT}\mapsto X$, which shall be less non-linear. In the following, such an auxiliary transform is derived for the case of Gaussian, lognormal, Gumbel, and Weibull parametric p-boxes.

\textbf{\emph{Gaussian p-box.}}
Consider a Gaussian parametric p-box where the mean value $\mu_X$ and standard deviation $\sigma_X$ are given in intervals. Instead of using the uniform distribution $C$, a standard Gaussian variable $\xi\sim\mathcal{N}(0,\,1)$ can be used in the independent vector $\vV=\prt{\xi,\mu_X,\sigma_X}$. Then, the isoprobabilistic transform reads:
\begin{equation} \label{eq:gaussian}
X = \mu_X + \sigma_X\cdot \xi. 
\end{equation}
This transform is simpler than the inverse CDF formulation in Eq.~(\ref{eq:augmentedmodel:separate}) because the unbounded variable $X$ is modelled by means of the unbounded variable $\xi$. In fact, the transform features a linear mapping from $X$ to $\xi$. 

\textbf{\emph{Lognormal p-box.}}
Considering a lognormal parametric p-box with interval-valued parameters $\prt{\lambda_X,\zeta_X}$, such that $\log X\sim \mathcal{N}\prt{\lambda_X,\zeta_X}$, then: 
\begin{equation}
X = e^{\lambda_X + \zeta_X\cdot \xi},
\end{equation}
where $\xi$ follows a standard Gaussian distribution, and $\prt{\lambda_X,\,\zeta_X}$ are obtained by the usual equations from the mean value $\mu_X$ and standard deviation $\sigma_X$. Note that a lognormal p-box defined by lower and upper bounds on mean $\mu_X$ and standard deviation $\sigma_X$ leads to a non-rectangular domain in the $\prt{\lambda_X,\,\zeta_X}$-plane. This is, however, not an issue since the experimental designs will be sampled in the $\prt{\mu_X,\,\sigma_X}$-space.  

\textbf{\emph{Gumbel p-box.}}
The CDF of a Gumbel distribution is defined as: 
\begin{equation}
F_{\mathcal{GU}}(x|\alpha,\beta) = \exp\bra{-\exp\prt{-(x-\alpha)/\beta}},
\end{equation}
where $\alpha = \mu_X - \beta\gamma_e$ and $\beta=\sigma_X\sqrt{6}/\pi$ are the distribution hyper-parameters ($\mu_X$ and $\sigma_X$ denote the mean and standard deviation) and $\gamma_e=0.5772\ldots$ is the Euler-Mascheroni constant. Then, a standard Gumbel variable is denoted by $\varpi\sim\mathcal{GU}(\alpha=0,\beta=1)$. It follows that a distribution with arbitrary hyper-parameter values $\prt{\alpha,\beta}$ can be formulated in terms of a standard Gumbel distribution as follows: 
\begin{equation}
F_{\mathcal{GU}}(x|\alpha,\beta) = F_{\mathcal{GU}}(\varpi|0,1),
\end{equation}
which reduces to:
\begin{equation}
(x-\alpha)/\beta = \varpi.
\end{equation}
Hence, a standard Gumbel distribution can be transformed into an arbitrary Gumbel distribution by:
\begin{equation}
X = \alpha + \beta\cdot \varpi,
\end{equation}
which resembles the construction for Gaussian variables in Eq.~(\ref{eq:gaussian}) with different transformation parameters. Accordingly, a Gumbel p-box can be represented by three random variables, including two uniform and one standard Gumbel.

\textbf{\emph{Weibull p-box.}}
The CDF of a Weibull distribution is defined as:
\begin{equation}
F_{\mathcal{W}}(x|\alpha,\beta) = 1-\exp\bra{-\prt{x/\alpha}^\beta},
\end{equation}
where $x\in\bra{0,\,\infty}$ and $\prt{\alpha,\beta}$ are its hyper-parameters. Then, a standard Weibull distribution is denoted by $\varpi\sim\mathcal{W}(\alpha=1,\beta=1)$, which also corresponds to a standardized exponential distribution. It follows that a distribution with arbitrary hyper-parameter values $\prt{\alpha,\beta}$ can be formulated in terms of an exponential distribution as follows: 
\begin{equation}
X = \alpha\cdot \varpi^{1/\beta}.
\end{equation}
Accordingly, a Weibull p-box can be represented by three random variables including two uniform and one exponential. 

\subsubsection{Bounded distributions}
\begin{figure}[!ht]
\centering
\includegraphics[width=0.45\linewidth]{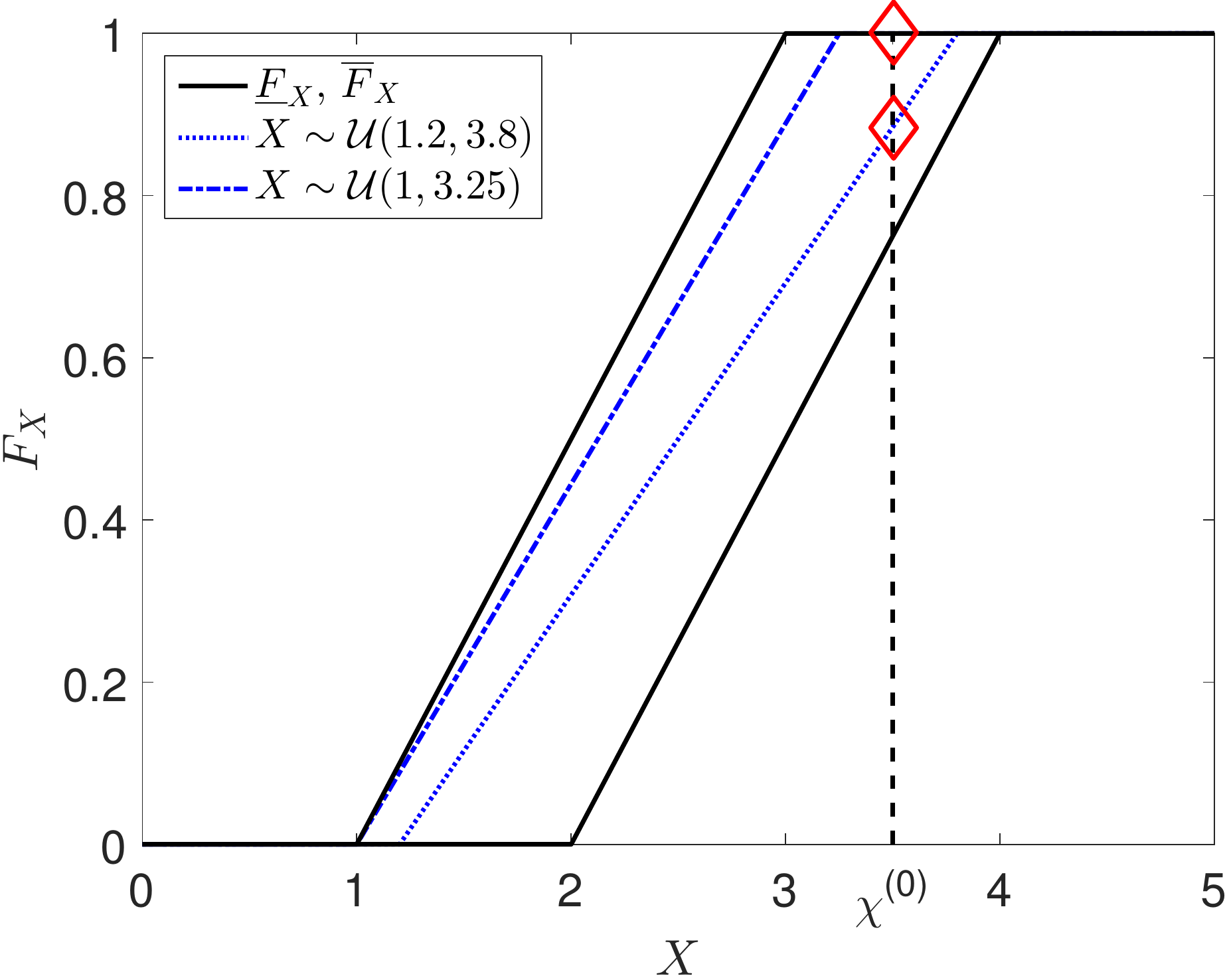}
\caption{\label{fig:bounds} Illustration of phantom points for bounded distributions. $X\sim\cU(a,b)$, where $a\in[1,\,2]$ and $b\in[3,\,4]$.}
\end{figure}

%
When dealing with distributions that are bounded on two sides, \ie $X\in\bra{\underline{x},\,\overline{x}}$, the phantom points cannot be generated by the approach detailed in the previous section. The problem is illustrated in Figure~\ref{fig:bounds}, which shows a parametric p-box consisting of a uniform distribution with interval-valued distribution bounds, \ie $X\sim\cU\prt{a,b}$, where $a\in[1,\,2]$ and $b\in[3,\,4]$. Consider the experimental design point $\chi^{(0)}=3.5$ and the two realizations with $\acc{a=1.2,\, b=3.8}$ and $\acc{a=1, \, b=3.25}$. For the first realization, the phantom point for $\chi^{(0)}$ reads:
$$\vvf^{(0)(1)} = \acc{a=1.2, \, b=3.8, \, c=F_{\cU}\prt{\chi^{(0)}\middle| 1.2, 3.8}\approx 0.885}.$$
For the second realization, however, the phantom point lies outside of the support interval $X\in[1, \, 3.25]$. Hence, the construction of a phantom point is not meaningful in this case. For $\chi^{(0)}=3.5$ in particular, a phantom point is only feasible when $a\in\bra{1,\,2}$ and $b \geq 3.5$ at the same time. 

In other words, the generation of phantom points is submitted to constraints in case of bounded parametric distributions. In order to generate phantom points efficiently, these constraints must be taken into account. Therefore, Eq.~(\ref{eq:nph}) is transformed into:
\begin{equation} \label{eq:nphb}
\mathfrak{c}_i^{(j)(k)} = F_{X_i} \prt{ \chi_i^{(j)}\middle| \vtau_i^{(j)(k)}} \qquad \text{s.t.} \qquad \chi_i^{(j)}\in\bra{\underline{x}_{\vtau_i^{(j)(k)}},\ \overline{x}_{\vtau_i^{(j)(k)}}},
\end{equation}
where $X_{\vtau_i^{(j)(k)}}$ denotes a conditional distribution defined within the bounds $X_{\vtau_i^{(j)(k)}}\in\bra{\underline{x}_{\vtau_i^{(j)(k)}},\ \overline{x}_{\vtau_i^{(j)(k)}}}$. When $\chi_i^{(j)}$ is outside the interval bounds, no phantom point is generated for the realization $\vtau_i^{(j)(k)}$.

\section{Imprecise Sobol' indices by PCE} \label{sec:isp}
\subsection{Reordering}
%
When an augmented PCE model is available as in Eq.~(\ref{eq:augpce}), the imprecise Sobol' indices can be obtained by post-processing this augmented PCE model. The imprecise Sobol' indices shall be depending on $\vC$ and conditioned on $\vT$. In order to do so, the augmented PCE model is recast as follows: 
\begin{equation}
W\approx \cm^{(\text{P})}\prt{\vC,\vT} = \sum_{\ua\in\cA} a_{\ua}\, \psi_{\ua_{\vC}}\prt{\vC} \, \psi_{\ua_{\vT}}\prt{\vT},
\end{equation}
where $\ua=\prt{\ua_{\vC},\ua_{\vT}}$ forms the same index set as in Eq.~(\ref{eq:augpce}) and $\psi_{\ua}\prt{\vV} = \psi_{\ua_{\vC}}\prt{\vC}\cdot \psi_{\ua_{\vT}}\prt{\vT}$ are the corresponding multivariate orthonormal polynomials. With this notation, we emphasize that any polynomial of $\prt{\vC,\vT}$ can be cast as a product of two polynomials of $\vC$ and $\vT$, respectively, due to the initial tensor product construction. Then, the computational model can be further rearranged, for a given value $\vt$ as a function of $\vC$ only:
\begin{equation} \label{eq:wttt}
W\prt{\vt}\approx \cm^{(\text{P})}\prt{\vC,\vt}=\sum_{\ua\in\cA} \prt{a_{\ua} \psi_{\ua_{\vT}}\prt{\vt}}\ \psi_{\ua_{\vC}}\prt{\vC} 
= \sum_{\ua\in\cA} a_{\ua,\vt}(\vt) \ \psi_{\ua_{\vC}}\prt{\vC}, 
\end{equation}
where $a_{\ua,\vt}(\vt)=a_{\ua}\ \psi_{\ua_{\vT}}\prt{\vt}$ is a new coefficient dependent on the value of $\vt$. Note that several $\ua_{\vT}$ may correspond to the same $\ua_{\vC}$ in $\ua$. Hence, Eq.~(\ref{eq:wttt}) can be further rewritten to a model with proper variance decomposition as in Eq.~(\ref{eq:pceso}):
\begin{equation}
W\prt{\vt}\approx \cm^{(\text{P})}\prt{\vC,\vt} 
= \sum_{\ua_{\vC}^*\in\cA_{\vC}^*} \prt{\sum_{i=1,\ldots,n_{\cA}} \Ii_{\acc{\ua_{\vC}^{(i)}=\ua_{\vC}^*}}\prt{\ua_{\vC}^{(i)}}\ a_{\ua^{(i)}}\ \psi_{\ua_{\vT}^{(i)}}\prt{\vt}  } \psi_{\ua_{\vC}^*}\prt{\vC},
\end{equation}
where $i$ marks the $1,\ldots,n_{\cA}$ multi-indices in $\cA$, $\Ii$ is the indicator function with $\Ii=1$ for $\ua_{\vC}^{(i)}=\ua_{\vC}^*$ and $\Ii=0$ otherwise, $\cA_{\vC}^*$ is the set of unique multi-indices $\ua_{\vC}$ in $\cA$, and $\ua^{(i)}=\prt{\ua_{\vC}^{(i)},\ua_{\vT}^{(i)}}$. Rearranging the terms to obtain the same structure as in Eq.~(\ref{eq:pceso}) leads to the following coefficients depending on $\vt$:
\begin{equation} \label{eq:aunique}
a_{\ua_{\vC}^*}\prt{\vt} = \sum_{i=1,\ldots,n_{\cA}} \Ii_{\acc{\ua_{\vC}^{(i)}=\ua_{\vC}^*}}\prt{\ua_{\vC}^{(i)}}\ a_{\ua^{(i)}}\ \psi_{\ua_{\vT}^{(i)}}\prt{\vt}.
\end{equation} 
The simplified decomposition then reads:
\begin{equation} \label{eq:augpceref}
W_{\vt}\approx \cm^{(\text{P})}\prt{\vC, \vT=\vt} = \sum_{\ua_{\vC}^*\in\cA_{\vC}^*} a_{\ua_{\vC}^*}(\vT=\vt)\  \psi_{\ua_{\vC}^*}\prt{\vC}. 
\end{equation}

\subsection{Optimization}
%
%
Finally, in order to obtain the extreme values of the Sobol' indices, the optimizations in Eq.~(\ref{eq:sobolminmax0}) use the reformulated model in Eq.~(\ref{eq:augpceref}). Then, the bounds of imprecise Sobol' indices can be computed by solving two optimization problems:
\begin{equation} \label{eq:imp}
\underline{S}^{(\text{P})}_{i_1\ldots i_s} = \min_{\vt\in\cD_{\vT}} \frac{D_{i_1\ldots i_s}(\vt)}{D(\vt)}= \min_{\vt\in\cD_{\vT}} \left[ \sum_{\ua^*_{\vC}\in\cI_{i_1\ldots i_s}} a^2_{\ua^*_{\vC}}\prt{\vt} \middle/ \sum_{\ua_{\vC}^*\in\cA^*_{\vC}, \ \ua_{\vC}^*\not= \ve{0}} a^2_{\ua^*_{\vC}}\prt{\vt}  \right],
\end{equation}
\begin{equation}\label{eq:imp2}
\overline{S}^{(\text{P})}_{i_1\ldots i_s} = \max_{\vt\in\cD_{\vT}} \frac{D_{i_1\ldots i_s}(\vt)}{D(\vt)}=\max_{\vt\in\cD_{\vT}} \left[ \sum_{\ua^*_{\vC}\in\cI_{i_1\ldots i_s}} a^2_{\ua^*_{\vC}}\prt{\vt} \middle/ \sum_{\ua_{\vC}^*\in\cA^*_{\vC}, \ \ua_{\vC}^*\not= \ve{0}} a^2_{\ua^*_{\vC}}\prt{\vt}  \right],
\end{equation}
In a nutshell, once the augmented PCE has been calibrated, the conditional Sobol' indices $S_{i_1\ldots i_s}\prt{\vt}$ are simply \emph{ratios} of multivariate orthogonal polynomials in $\vt$. The two optimization problems in Eqs.~(\ref{eq:imp}) and (\ref{eq:imp2}) can then be solved at reasonable cost through global optimization algorithms, such as genetic or simulated annealing algorithms. 

\subsection{Connection to Bayesian hierarchical models}
The imprecise Sobol' indices computed here for parametric p-boxes can be extended to Bayesian hierarchical models with little modifications. When interpreting $\vt$ as distributions rather than interval-valued quantities, Eqs.~(\ref{eq:imp}) and (\ref{eq:imp2}) converge to a single equation describing a distribution of Sobol' indices:
\begin{equation} \label{eq:bayes}
S^{(\text{P})}_{i_1\ldots i_s}\prt{\vT} = \frac{D_{i_1\ldots i_s}(\vT)}{D(\vT)}=  \left. \sum_{\ua^*_{\vC}\in\cI_{i_1\ldots i_s}} a^2_{\ua^*_{\vC}}\prt{\vT} \middle/ \sum_{\ua_{\vC}^*\in\cA^*_{\vC}, \ \ua_{\vC}^*\not= \ve{0}} a^2_{\ua^*_{\vC}}\prt{\vT}  \right..
\end{equation}
As a direct consequence, the \emph{distribution} of any Sobol' index $S_{i_1\ldots i_s}^{(\text{P})}$ due to the epistemic uncertainty in $\vt$ could be obtained at low cost by Monte Carlo sampling of Eq.~(\ref{eq:bayes}).

\section{Applications} \label{sec:appl}
\subsection{Step-by-step illustration of algorithm}
%
%
\subsubsection{Problem statement}
A simple two-dimensional function is considered here to visualize the algorithm for computing the imprecise Sobol' indices. The analytical function is defined as:
\begin{equation} \label{eq:f8}
f_1(\vx) = x_1\cdot x_2,
\end{equation}
where $x_1$ and $x_2$ are modelled by parametric p-boxes. The input variables are modelled by Gaussian distributions with interval-valued mean and standard deviation, namely ${\mu_i\in\bra{-1,1}}$ and ${\sigma_i\in\bra{0.5,1.0}}$. As seen in the sequel, this function allows us to derive analytically the conditional Sobol' indices as well as their bounds. 

\subsubsection{Imprecise Sobol' indices}
Each input parameter may be cast as $X_i=\mu_i+\sigma_i\cdot \xi_i$ where $\xi_i$ is a standard normal variable. Then, the augmented computational model can be written as:
\begin{multline} \label{eq:f8aug}
f^{(\text{aug})}_1\prt{\ve{\mu},\ve{\sigma},\ve{\xi}} = \prt{\mu_1+\sigma_1\cdot \xi_1}\cdot \prt{\mu_2+\sigma_2\cdot \xi_2} \\
=\prt{\mu_1\cdot \mu_2} + \prt{\mu_2\cdot \sigma_1}\cdot \xi_1 + \prt{\mu_1\cdot\sigma_2}\cdot \xi_2 + \prt{\sigma_1\cdot\sigma_2}\cdot \xi_1\cdot\xi_2.
\end{multline} 
The variance of the response variable conditioned on the values of $\ve{\mu}$ and $\ve{\sigma}$ then reads: 
\begin{equation}
D\prt{\ve{\mu},\ve{\sigma}} \equiv \var{f_1|\ve{\mu},\ve{\sigma}} = \prt{\mu_2\sigma_1}^2 + \prt{\mu_1\sigma_2}^2 + \prt{\sigma_1\sigma_2}^2.
\end{equation}
Hence, the first order Sobol' indices can be computed as a function of $\vt=\prt{\ve{\mu}, \ve{\sigma}}$:
\begin{equation}
S_1\prt{\vt} = \frac{\prt{\mu_2\sigma_1}^2}{\prt{\mu_2\sigma_1}^2 + \prt{\mu_1\sigma_2}^2 + \prt{\sigma_1\sigma_2}^2}, \qquad S_2\prt{\vt} = \frac{\prt{\mu_1\sigma_2}^2}{\prt{\mu_2\sigma_1}^2 + \prt{\mu_1\sigma_2}^2 + \prt{\sigma_1\sigma_2}^2},
\end{equation}

The analytical derivation of the extreme Sobol' indices results in the following bounds for the first order Sobol' indices:
\begin{equation} \label{eq:f8sobol1}
\underline{S}_1 = \underline{S}_2 = 0.0, \qquad \overline{S}_1=\overline{S}_2=0.8.
\end{equation} 
As an example, the minimal first order Sobol' index for $X_1$ is obtained by setting $\mu_2=0$. 
Similarly to the first order indices, the total order Sobol' indices can be written as:
\begin{equation}
S_1^{(T)}\prt{\vt} = \frac{\prt{\mu_2\sigma_1}^2 + \prt{\sigma_1\sigma_2}^2}{\prt{\mu_2\sigma_1}^2 + \prt{\mu_1\sigma_2}^2 + \prt{\sigma_1\sigma_2}^2}, \qquad S_2^{(T)}\prt{\vt} = \frac{\prt{\mu_1\sigma_2}^2 + \prt{\sigma_1\sigma_2}^2}{\prt{\mu_2\sigma_1}^2 + \prt{\mu_1\sigma_2}^2 + \prt{\sigma_1\sigma_2}^2},
\end{equation}
The bounds of the total order Sobol' indices read:
\begin{equation} \label{eq:f8sobolt}
\underline{S}_1^{(T)} = \underline{S}_2^{(T)} = 0.2, \qquad \overline{S}_1^{(T)}=\overline{S}_2^{(T)} = 1.0.
\end{equation}
As an example, the minimal total order Sobol' index for $X_1$ is obtained by setting $\mu_1 = \pm 1$, $\sigma_1=0.5$, $\mu_2=0$, and $\sigma_2=1$. Due to the symmetry of the problem definition, \ie computational model $f_1^{(\text{aug})}$ and input p-boxes $X_i$, the Sobol' indices are symmetrical in $\prt{\mu_i,\sigma_i}$. Note that in this example, the first and total order Sobol' indices are not identical due to the interactive term $\prt{\sigma_1\cdot\sigma_2}\cdot \xi_1\cdot\xi_2$ in Eq.~(\ref{eq:f8aug}). 

\subsubsection{Augmented PCE-based Sobol' indices}
Using the \textsc{Matlab}-based uncertainty quantification framework {\sc UQLab} \citep{MarelliICVRAM2014,UQdoc_09_104}, the augmented PCE of $f_1^{(\text{aug})}$ and the parametric p-boxes results in the following set of multi-indices $\cA$:
\begin{equation}
\cA =  \prt{
\begin{array}{c}
\ua^{(1)} \\ 
\vdots \\ 
\ua^{(10)}
\end{array} 
} = \prt{
\begin{array}{c c c c c c}
0   &  0    & 0    & 0    & 0   &  0 \\
    0   &  0 &    1 &    0 &    0 &    1 \\
     0   &  0 &    1 &    1 &    0 &    0 \\
     1    & 0  &   0  &   0  &   0  &   1\\
     1     & 0  &   0  &   1  &   0  &   0\\
     0&     0    & 1    & 0    & 1    & 1\\
     0 &    1 &    1&     0 &    0 &    1\\
     0  &   1  &   1 &    1  &   0  &   0\\
     1   &  0   &  0  &   0   &  1   &  1\\
     0    & 1    & 1   &  0    & 1    & 1
\end{array}
},
\end{equation}
which contains ten vectors $\ua$ and where the indices in each row correspond to the augmented input vector $\prt{\mu_1, \sigma_1, \xi_1, \mu_2, \sigma_2, \xi_2}$. Splitting the multi-index set $\cA$ to the contributions of the epistemic ($\mu_1,\,\sigma_1,\,\mu_2,\,\sigma_2$) and aleatory ($\xi_1,\,\xi_2$) variables leads to:
\begin{equation}
\cA_{\vT} = \prt{
\begin{array}{c c c c}
 0    & 0    & 0   &  0 \\
0   &  0 &       0 &    0  \\
0   &  0 &        1 &    0  \\
1    & 0  &     0  &   0  \\
1     & 0  &      1  &   0  \\
0&     0    &  0    & 1    \\
0 &    1 &         0 &    0 \\
0  &   1  &       1  &   0 \\
1   &  0   &     0   &  1   \\
0    & 1    &  0    & 1    
\end{array}
}, \qquad 
\cA_{\vC} = \prt{
\begin{array}{c c c c c c}
 0       &  0 \\
      1 &     1 \\
       1 &       0 \\
      0  &      1\\
      0  &    0\\
   1       & 1\\
      1&     1\\
     1 &     0\\
    0  &    1\\
     1     & 1
\end{array}
}.
\end{equation}
Note that according to Eq.~(\ref{eq:f8aug}), only four elements should be part of $\cA$. However, due to the normalization to a standard space (uniform distributions to $[-1,1]$ and Gaussian to standard normal space), an additional six elements are present.

$\cA_{\vC}^*$ summarizes the set of unique row vectors $\ua_{\vC}$ in $\cA_{\vC}$ and contains the following four elements:
\begin{equation}
\cA_{\vC}^* = \prt{
\begin{array}{c}
\ua_{\vC}^{*(1)} \\
\ua_{\vC}^{*(2)} \\
\ua_{\vC}^{*(3)} \\
\ua_{\vC}^{*(4)} 
\end{array}
}
=
\prt{ \begin{array}{cc}
0 & 0 \\
1 & 1 \\
1 & 0 \\
0 & 1
\end{array} }.
\end{equation}
The final coefficients are composed as follows (see also Eq.~(\ref{eq:aunique})):
\begin{eqnarray}
a_{\ua_{\vC}^{*(1)}}\prt{\vt} &=& a_{\ua^{(1)}}\psi_{\ua_{\vT}^{(1)}}(\vt) + a_{\ua^{(5)}}\psi_{\ua_{\vT}^{(5)}}(\vt), \label{eq:aalpha:1}\\ 
a_{\ua_{\vC}^{*(2)}}\prt{\vt} &=& a_{\ua^{(2)}}\psi_{\ua_{\vT}^{(2)}}(\vt) + a_{\ua^{(6)}}\psi_{\ua_{\vT}^{(6)}}(\vt) + a_{\ua^{(7)}}\psi_{\ua_{\vT}^{(7)}}(\vt) + a_{\ua^{(10)}}\psi_{\ua_{\vT}^{(10)}}(\vt), \label{eq:aalpha:2}\\
a_{\ua_{\vC}^{*(3)}}\prt{\vt} &=& a_{\ua^{(3)}}\psi_{\ua_{\vT}^{(3)}}(\vt) + a_{\ua^{(8)}}\psi_{\ua_{\vT}^{(8)}}(\vt),\\
a_{\ua_{\vC}^{*(4)}}\prt{\vt} &=& a_{\ua^{(4)}}\psi_{\ua_{\vT}^{(4)}}(\vt) + a_{\ua^{(9)}}\psi_{\ua_{\vT}^{(9)}}(\vt),
\end{eqnarray}
where Eqs.~(\ref{eq:aalpha:1}) and~(\ref{eq:aalpha:2}) can be simplified to:
\begin{eqnarray}
a_{\ua_{\vC}^{*(1)}}\prt{\vt} &=& a_{\ua^{(1)}}+ a_{\ua^{(5)}}\psi_{\ua_{\vT}^{(5)}}(\vt),\\ 
a_{\ua_{\vC}^{*(2)}}\prt{\vt} &=& a_{\ua^{(2)}} + a_{\ua^{(6)}}\psi_{\ua_{\vT}^{(6)}}(\vt) + a_{\ua^{(7)}}\psi_{\ua_{\vT}^{(7)}}(\vt) + a_{\ua^{(10)}}\psi_{\ua_{\vT}^{(10)}}(\vt).
\end{eqnarray}
The variance decomposition of the PCE model then reads (see also Eq.~(\ref{eq:augpceref})):
\begin{equation}
\cm^{(\text{P})}\prt{\vC,\vT} = \sum_{i=1,\ldots,4} a_{\ua_{\vC}^{*(i)}}\prt{\vT} \ \psi_{\ua_{\vC}^{(i)}}\prt{\vC},
\end{equation}
where $\vC = \prt{\xi_1,\xi_2}$. By optimization on $\vt$, the PC-based imprecise Sobol' indices can be computed. In this particular case, we obtain the analytical results of Eqs.~(\ref{eq:f8sobol1}) and (\ref{eq:f8sobolt}) exactly. Indeed, the initial model $f_1$ is polynomial, as well as the isoprobabilistic transforms. In the end, a 10-term PCE in the augmented space is exact.

\subsection{Single-degree-of-freedom oscillator}
%
%
\subsubsection{Problem statement}
\begin{figure}[!ht]
\centering
\includegraphics[width=0.6\linewidth]{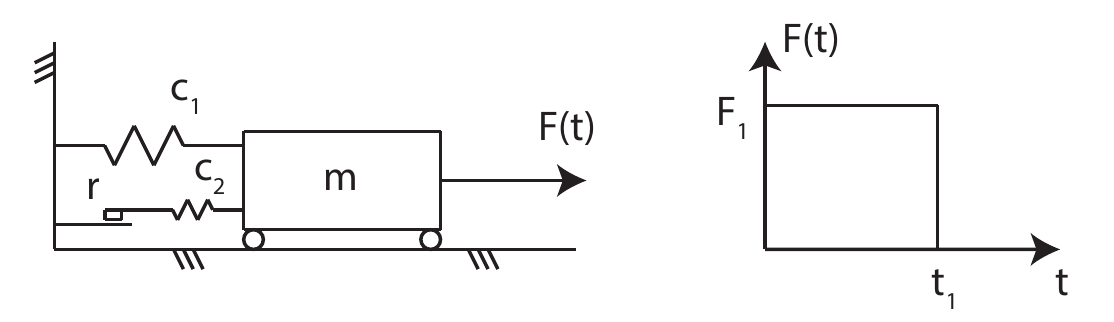}
\caption{\label{fig:osci:sketch} SDOF oscillator -- geometry sketch and definition of the variables}
\end{figure}

Consider the non-linear undamped single-degree-of-freedom (SDOF) oscillator sketched in Figure~\ref{fig:osci:sketch} \citep{Schueremans2005,Echard2011,Echard2013}. The corresponding computational model reads:
\begin{equation} \label{eq:sdof}
f_2\prt{r,F_1,t_1,c_1,c_2,m} = 3r - \left| \frac{2 F_1}{m\omega_0^2}\sin\prt{\frac{\omega_0T_1}{2}} \right|,
\end{equation}
where $m$ is the mass, $\acc{c_1, c_2}$ are the spring constants of the primary and secondary springs, $r$ is the displacement at which the secondary spring yields, $t_1$ is the duration of the loading, $F_1$ is the amplitude of the force and $\omega_0=\sqrt{\frac{c_1+c_2}{m}}$ is the natural frequency of the oscillator.

The input vector is modelled by a mix of probabilistic variables and parametric p-boxes accounting for the different levels of knowledge. The description of the input variables is provided in Table~\ref{tab:osci:variables}. It is assumed that the spring stiffnesses and the mass are well-known. Hence $\acc{c_1,c_2,m}$ are modelled by regular (``precise'') CDFs. On the other side, knowledge on $\acc{r,F_1,t_1}$ is supposed to be scarce. Hence, these variables are modelled by parametric p-boxes. As seen in Table~\ref{tab:osci:variables}, the parametric p-box is characterized by a distribution function with interval-valued mean value but constant standard deviation.  

\begin{table}[!ht]
\centering
\caption{SDOF oscillator -- definition of the parametric p-boxes in the input vector. \label{tab:osci:variables}}
\begin{tabular}{lllcc}
\hline
$X_i$ & Distribution && Mean value & Standard deviation\\
\hline
$r$ & Gaussian && $[0.49,0.51]$ & $0.05$ \\
$F_1$ & Gaussian && $[0.8,1.2]$& $0.2$  \\
$t_1$ & Gaussian && $[0.95,1.05]$& $0.2$ \\
$c_1$ & Gaussian && $1$& $0.1$ \\
$c_2$ & Gaussian && $0.1$& $0.01$\\
$m$ & Gaussian && $1$& $0.05$ \\
\hline
\end{tabular}
\end{table}

\subsubsection{Analysis}
In order to estimate the imprecise Sobol' indices, an augmented PCE model is trained for an augmented input space of dimension $M_{\text{aug}} = 6+3=9$. The number of model evaluations $N$ is varied to investigate the influence of the experimental design size. Note that the number of phantom points is set to $\nph=10$ for any $N$ in order to take full advantage of the phantom points. The sparse PCE is built with hyperbolic index sets ($q=0.75$) and degree-adaptive LARS with a maximal total polynomial degree $p=10$. The bounds of the imprecise Sobol' indices are then obtained through \textsc{Matlab}'s genetic optimization algorithm. 

A reference solution is obtained by using augmented PCE with a large sample size, \ie $N=1000$ and $\nph=10$. This results in an accurate meta-model and hence an accurate estimate of the imprecise Sobol' indices.

\subsubsection{Results}
The estimate for the first-order Sobol' indices $S_i$ are summarized in Table~\ref{tab:sdof:results}. The imprecise Sobol' indices are shown for the cases of $N=\acc{30,\,50,\,100,\,200}$. The larger the experimental design, the closer the estimates of bounds of the Sobol' indices to the reference values. In fact, an experimental design of $N=50$ is sufficient to obtain results with sufficient accuracy in this example. 

\begin{table}[!ht]
\centering
\caption{\label{tab:sdof:results} SDOF oscillator -- imprecise first-order Sobol' indices estimated by augmented PCE ($\nph=10$).}
\begin{tabular}{lcllll}
\hline
 & Reference & \multicolumn{4}{c}{Augmented PCE} \\
 &      & $N=30$ & $N=50$ & $N=100$ & $N=200$ \\
\hline
$S_r$ & $\bra{0.220,\,0.307}$& $\bra{0.312,\,0.313}$ & $\bra{0.223,\,0.305}$ & $\bra{0.220,\,0.307}$& $\bra{0.220,\,0.307}$\\
$S_{F_1}$ & $\bra{0.308,\, 0.459}$ & $\bra{0.373,\, 0.375}$ & $\bra{0.306,\,0.450}$ & $\bra{0.308,\,0.459}$ & $\bra{0.308,\,0.460}$\\
$S_{t_1}$ & $\bra{0.215,\,0.413}$& $\bra{0.214,\,0.215}$ & $\bra{0.224,\,0.415}$& $\bra{0.216,\,0.414}$ & $\bra{0.215,\,0.413}$\\
$S_{c_1}$ & $\bra{0.017,\,0.034}$ & $\bra{0.030,\ 0.030}$& $\bra{0.017,\,0.032}$&$\bra{0.017,\,0.034}$ & $\bra{0.017,\,0.034}$\\
$S_{c_2}$ & $\bra{0.000,\,0.000}$ & $\bra{0.001,\,0.001}$& $\bra{0.000,\,0.001}$ & $\bra{0.000,\, 0.000}$ & $\bra{0.000,\,0.000}$ \\
$S_{m}$ &   $\bra{0.003,\,0.006}$& $\bra{0.008,\ 0.009}$& $\bra{0.003,\,0.006}$& $\bra{0.003,\,0.006}$ & $\bra{0.003,\,0.006}$\\
&&&&&\\
$\widehat{err}_{gen}$ &  & $1.05\cdot 10^{-1}$ & $2.39\cdot 10^{-4}$ & $1.63\cdot 10^{-5}$ & $1.65\cdot 10^{-6}$ \\
\hline
\end{tabular}
\end{table}

In order to estimate the accuracy of the augmented PCE models, the relative generalization error is derived as the mean squared error between prediction and exact response value relative to the variance of the response variable. Based on a large validation set of $n=10^6$ samples generated from the augmented input vector $\vV$, the error estimate reads:
\begin{equation}\label{eq:errgen}
\widehat{err}_{gen} = \frac{\sum_{i=1}^n \prt{w_i - w_i^{(\text{P})}}^2}{\sum_{i=1}^{n}\prt{w_i-\mu_w}^2},
\end{equation}
where $\mathbb{W}=\acc{w_1,\ldots,w_n}$ are the exact response values corresponding to the validation set $\mathbb{V} = \acc{\vv_1,\ldots,\vv_n}$, $w_i^{(\text{P})}$ are the corresponding values predicted with the augmented PCE model, and $\mu_w=\frac{1}{n}\sum_{i=1}^{n} w_i$ is the mean value of the samples in $\mathbb{W}$. 

The values of the relative generalization error are listed in Table~\ref{tab:sdof:results}. As indicated for the bounds of the imprecise Sobol' indices, the augmented PCE model is accurate for large experimental designs. Values smaller than $\widehat{err}_{gen}\approx 10^{-4}$ indicate a high prediction accuracy.

\begin{figure}[!ht]
\centering
\subfigure[Bar plot \label{fig:sdof:results:bar}]{
\includegraphics[width=0.45\linewidth]{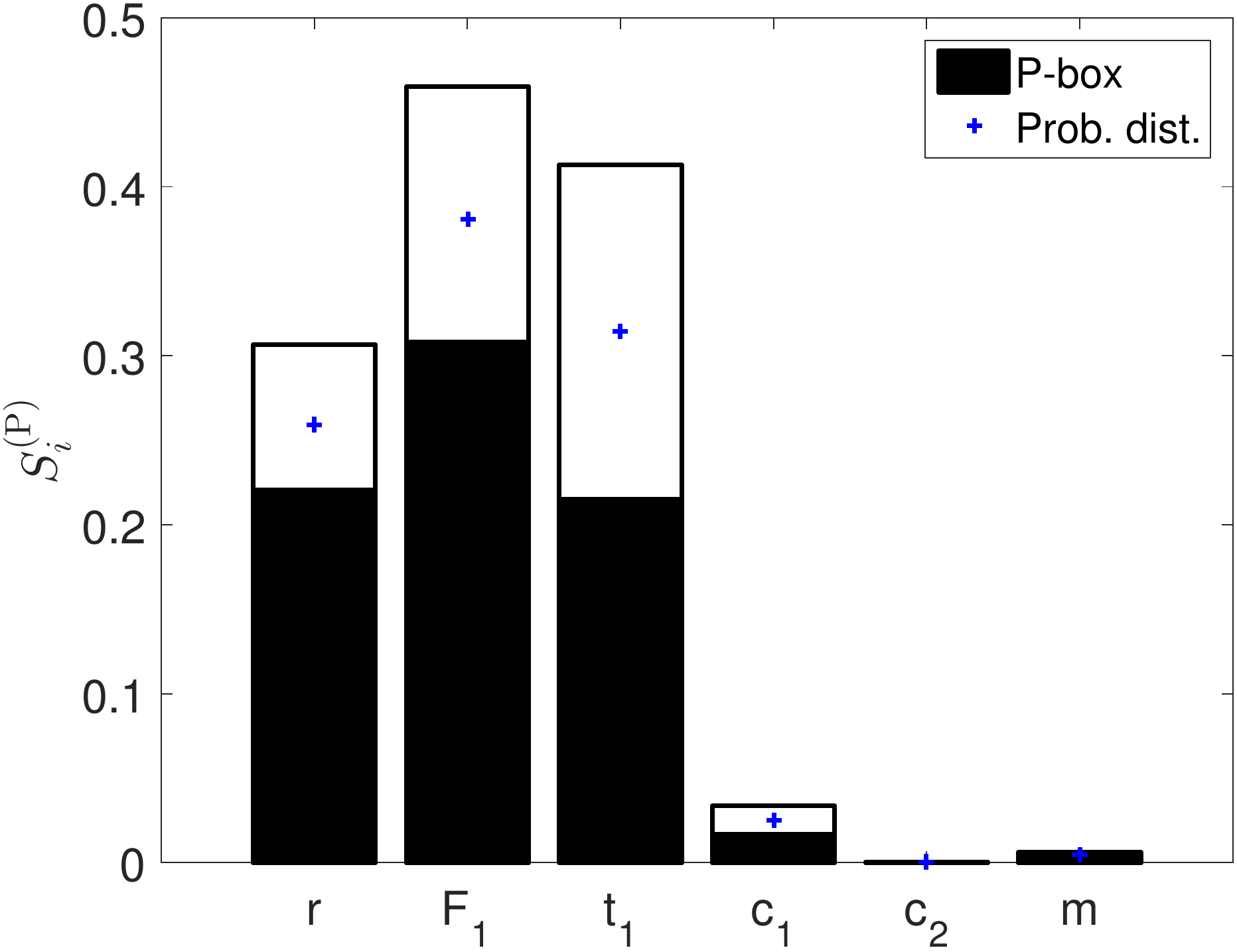}
}
\subfigure[Impact versus epistemic uncertainty plot \label{fig:sdof:results:xy}]{
\includegraphics[width=0.45\linewidth]{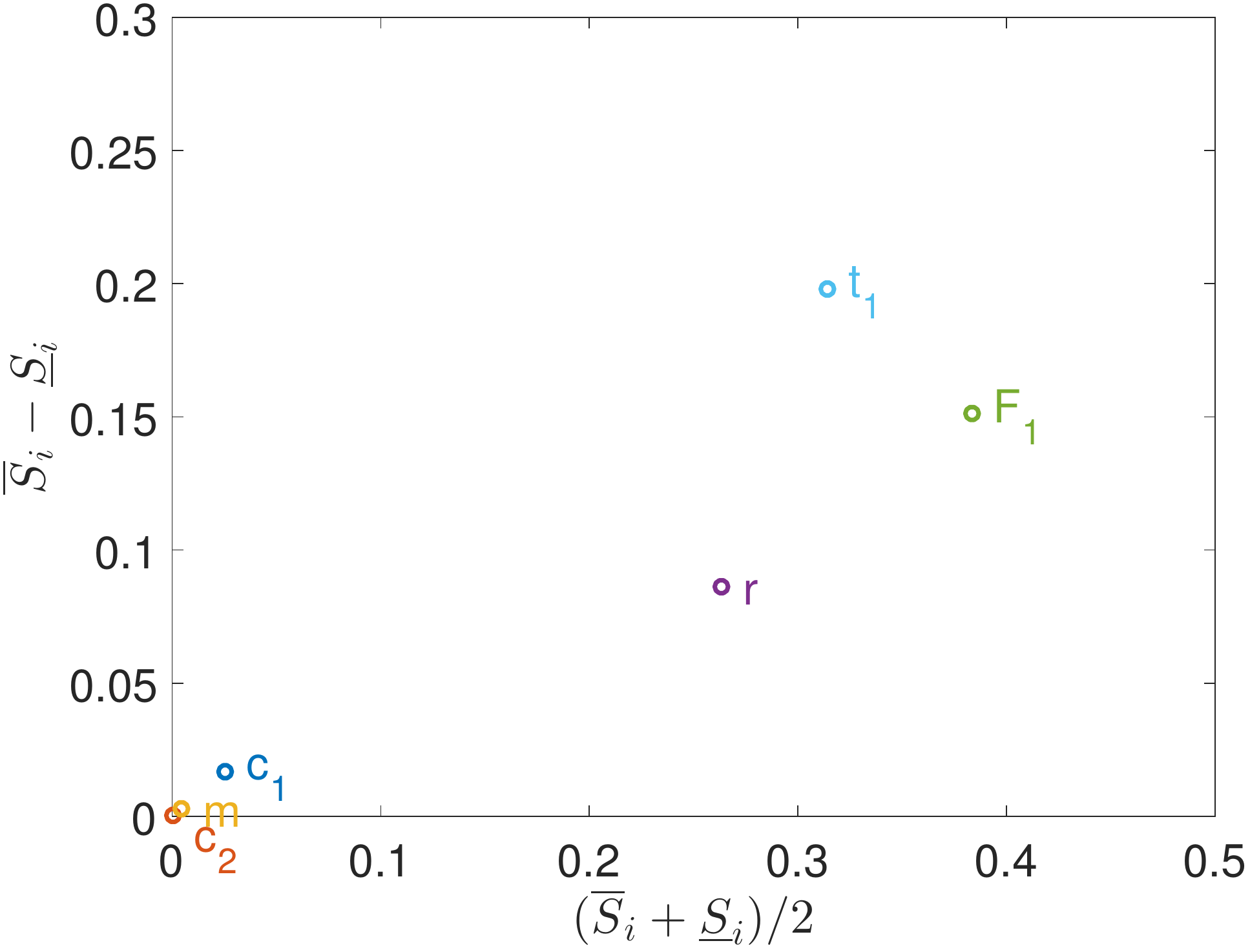}
}
\caption{\label{fig:sdof:results} SDOF oscillator -- imprecise Sobol' indices (first order) based on an augmented PCE model with $N=200$ and $\nph=10$. }
\end{figure}

Figure~\ref{fig:sdof:results:bar} illustrates the imprecise Sobol' indices obtained by $N=200$ and $\nph=10$ in a bar plot. The solid (black) part of the bars represents the lower bound of the first order Sobol' indices whereas the hollow bar represents the upper bound. For comparison to a purely probabilistic approach (\ie no epistemic uncertainty), the parametric p-boxes are ``pinched'' into classical probability distributions. The CDFs use the central value of the intervals as hyper-parameter: $r\sim\mathcal{N}\prt{0.50,\, 0.05}$, $F_1\sim\mathcal{N}\prt{1.0,\,0.2}$, and $t_1\sim\mathcal{N}\prt{1.00, \, 0.2}$. The resulting Sobol' indices are shown by blue markers in the bars. As expected, the imprecise Sobol' indices envelope the probabilistic estimates due to the fact that the ``probabilistic case'' is one possible realization of the parametric p-boxes.

A different visualization of the imprecise Sobol' indices is shown in Figure~\ref{fig:sdof:results:xy}. On the horizontal axis, the central value of the Sobol' indices is displayed, which is computed by $\prt{\overline{S}_i+\underline{S}_i}/2$. Hence, a variable placed on the right hand side of the plot has a large influence on the uncertainty in the response variable $Y$. On the vertical axis, the epistemic uncertainty of the Sobol' indices is shown defined as $\overline{S}_i-\underline{S}_i$. A variable placed at the upper end of the plot has a large epistemic uncertainty attached to the Sobol' indices. In this example, $F_1$ has the largest overall influence, whereas $t_1$ has the largest epistemic uncertainty in the estimate. Variables whose representative point lies close to the lower left corner are unimportant ($c_1$, $c_2$, and $m$ in the current example).

\subsection{Simply supported truss}

\subsubsection{Problem statement}
\begin{figure}[!ht]
\centering
\includegraphics[width=0.6\linewidth]{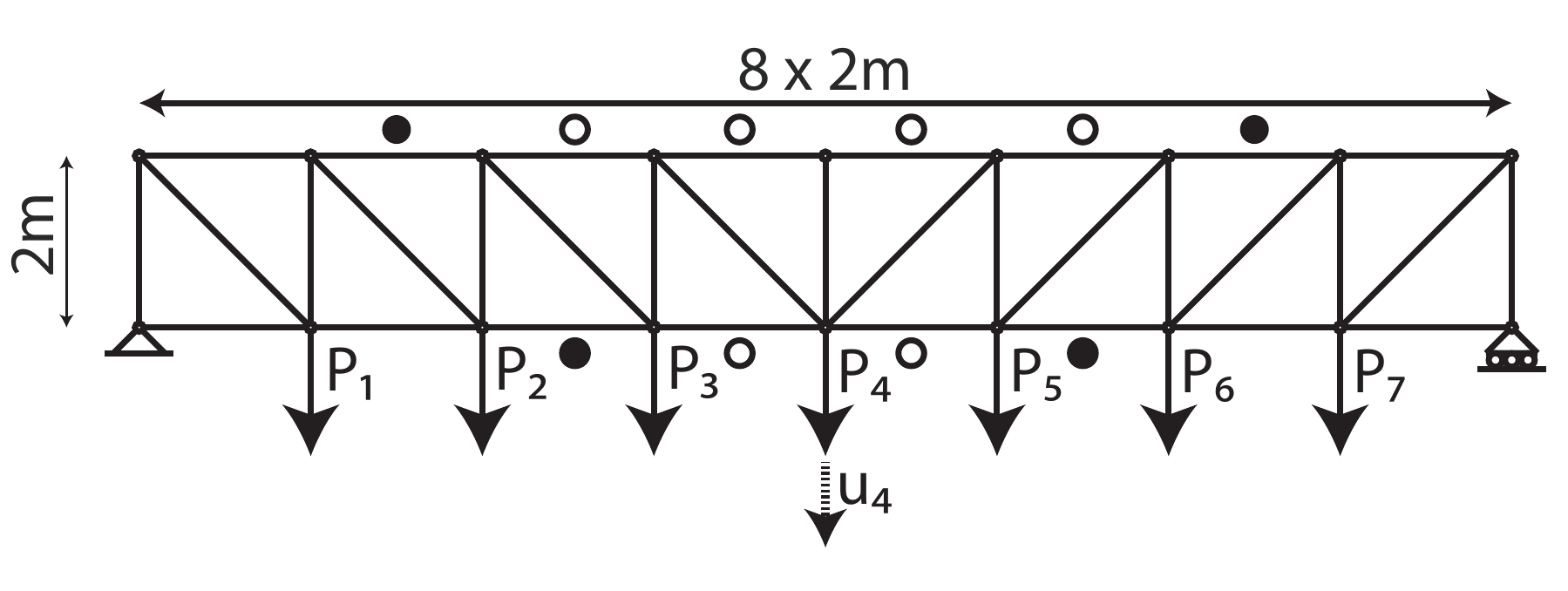}
\caption{\label{fig:truss:sketch} Truss structure -- sketch of the geometry and definition of the input variables.}
\end{figure}

Hurtado \cite{Hurtado2013} introduced the two-dimensional truss structure shown in Figure~\ref{fig:truss:sketch} for the purpose of reliability analysis. The truss is subjected to seven loads $P_i$ which are modelled with parametric p-boxes. The loads are defined by lognormal distributions with mean value $\mu_{P_i}\in[95,105]~{\rm kN}$ and standard deviation $\sigma_{P_i}\in[13,17]~{\rm kN}$. The geometry of the structure and the material properties are assumed constant. The modulus of elasticity is $E=200\cdot10^9~{\rm Pa}$, whereas the cross section area varies along the different bars: $A=0.00535~{\rm m}^2$ for bars marked by $\bullet$, $A=0.0068~{\rm m}^2$ for bars marked by $\circ$, and $A=0.004~{\rm m}^2$ for the remaining bars. The quantity of interest is the deflection at mid-span, denoted by $u_4$. 

\subsubsection{Analysis}
The deflection of the truss is computed by a finite element model (FEM), which is implemented in a \textsc{Matlab}-based in-house code. The FEM interprets each bar as a bar element, whereas the loads are modelled as point loads at the intersections of the corresponding bars as indicated in Figure~\ref{fig:truss:sketch}. 

In order to estimate the imprecise Sobol' indices, an augmented PCE model is computed based on an experimental design of $N=100$ samples generated by Latin-hypercube sampling in the unit-cube and subsequent isoprobabilistic transform to domain $\cD_{\vV}$. The original computational model requires a seven-dimensional input vector describing the loading forces. Due to the definition of the parametric p-boxes, the augmented PCE model requires a 21-dimensional input vector ($M_{\text{aug}}=21$). The number of phantom points $\nph$ is varied to examine the influence of these points onto the accuracy of the computed Sobol' indices. The bounds of the Sobol' indices are obtained by using \textsc{Matlab}'s built-in genetic algorithm with default settings as an optimization tool. 

The maximum deflection computed by the FE model corresponds to a monotone function of the load parameters. Hence, the extreme cases of the first order Sobol' indices can be obtained by setting one variable to $\acc{\mu_{P_i}=105~\text{kN}, \, \sigma_{P_i}=17~\text{kN}}$ and all others to $\acc{\mu_{P_i}=95~\text{kN}, \, \sigma_{P_i}=13~\text{kN}}$ or vice versa. The reference solution of the Sobol' indices are then obtained by PCE and a large experimental design. 

\subsubsection{Results}
\begin{figure} 
\centering
\subfigure[$\nph=1$ ($\widehat{err}_{gen}=3.4\cdot 10^{-1}$)]{
\includegraphics[width=0.45\linewidth]{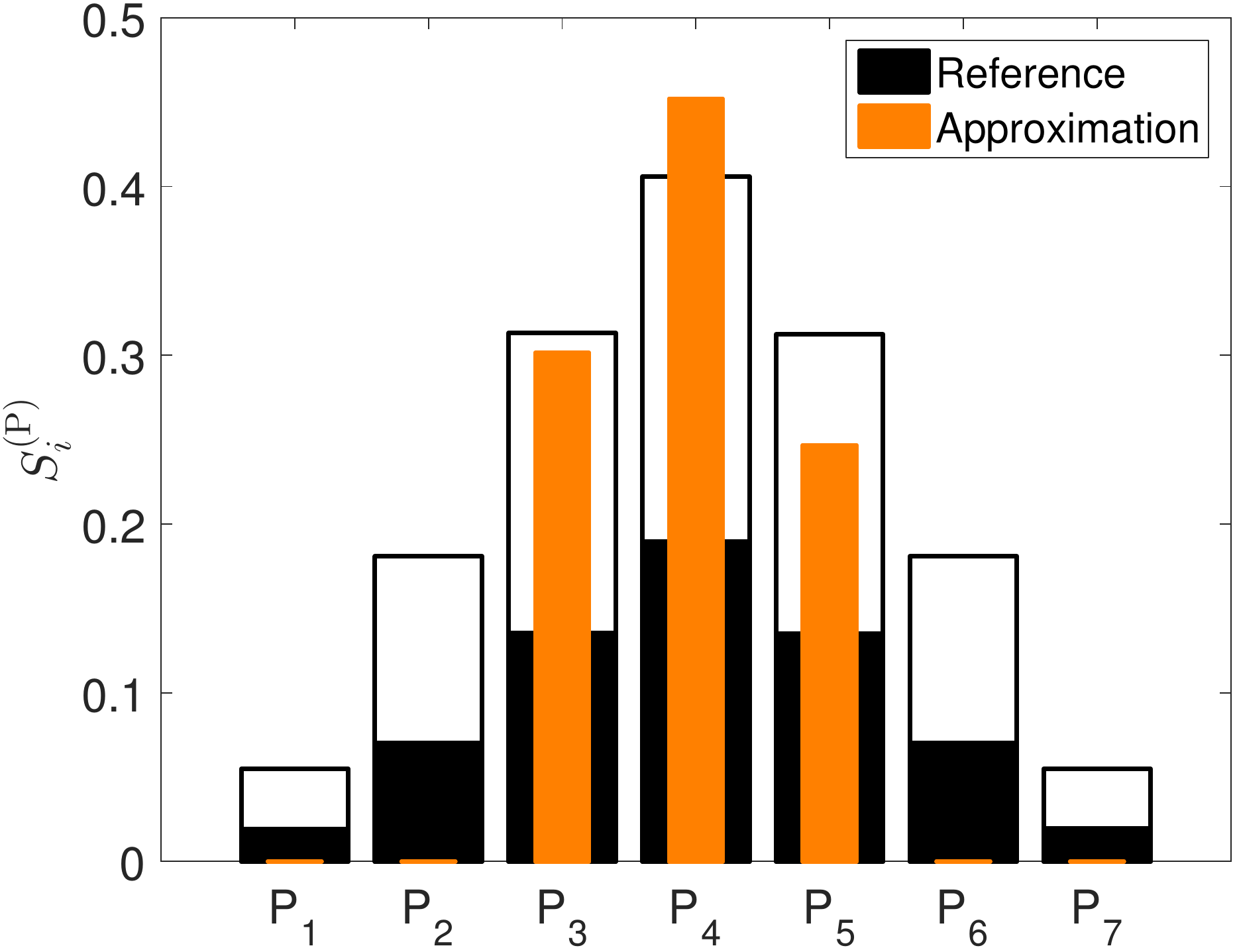}
}
\subfigure[$\nph=2$ ($\widehat{err}_{gen}=4.9\cdot 10^{-3}$)]{
\includegraphics[width=0.45\linewidth]{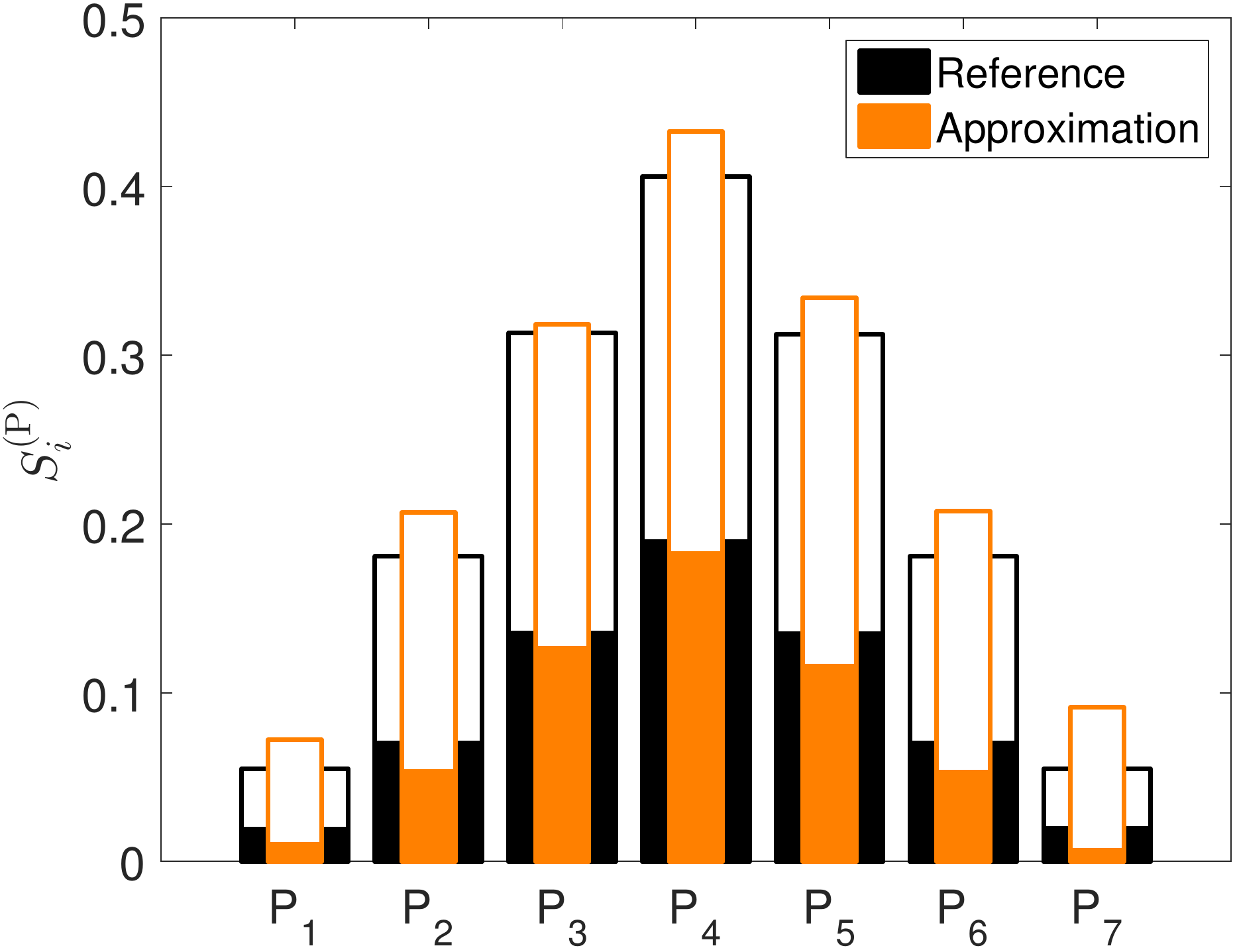}
}
\subfigure[$\nph=3$ ($\widehat{err}_{gen}=8.4\cdot 10^{-4}$)]{
\includegraphics[width=0.45\linewidth]{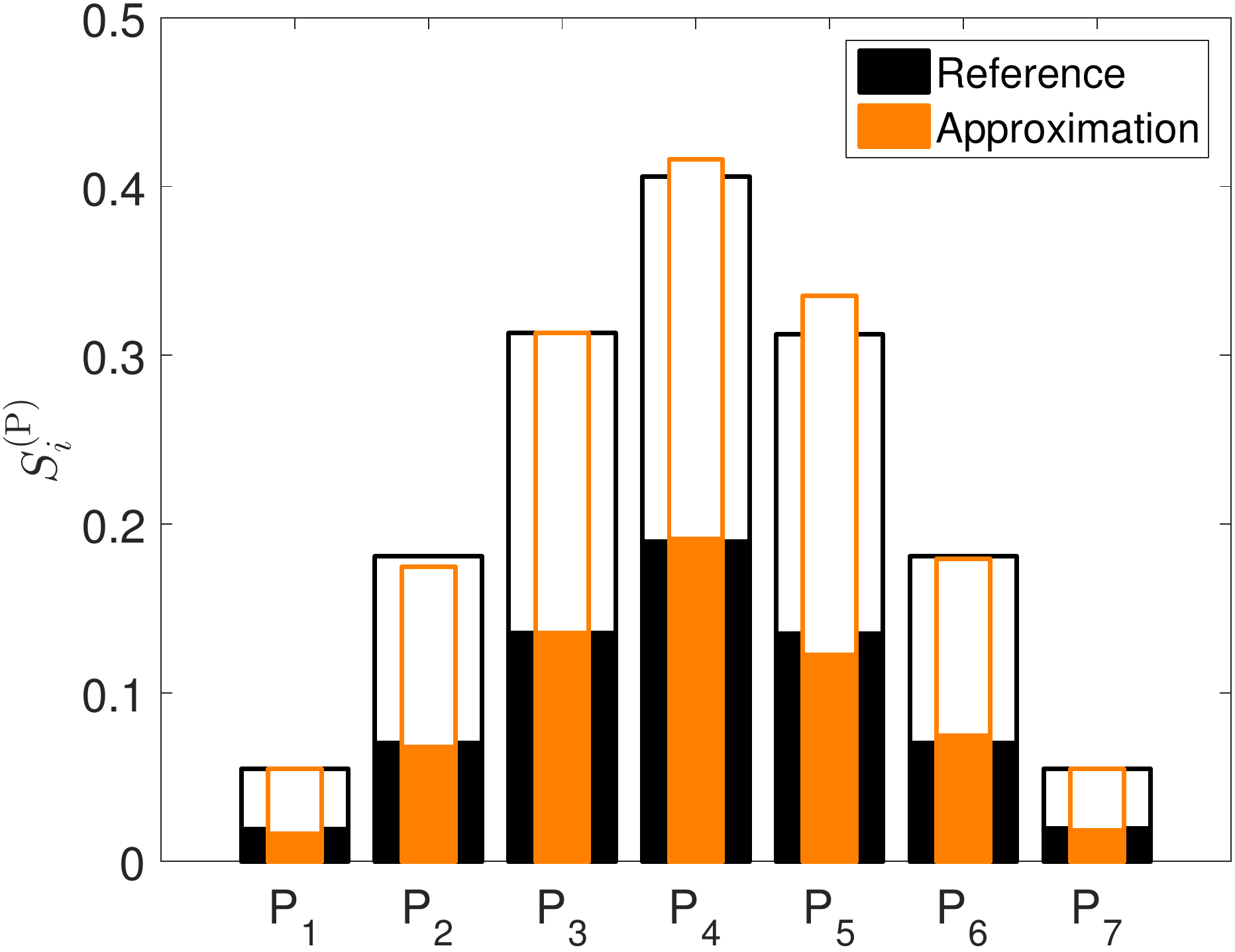}
}
\subfigure[$\nph=5$ ($\widehat{err}_{gen}=1.1\cdot 10^{-5}$)]{
\includegraphics[width=0.45\linewidth]{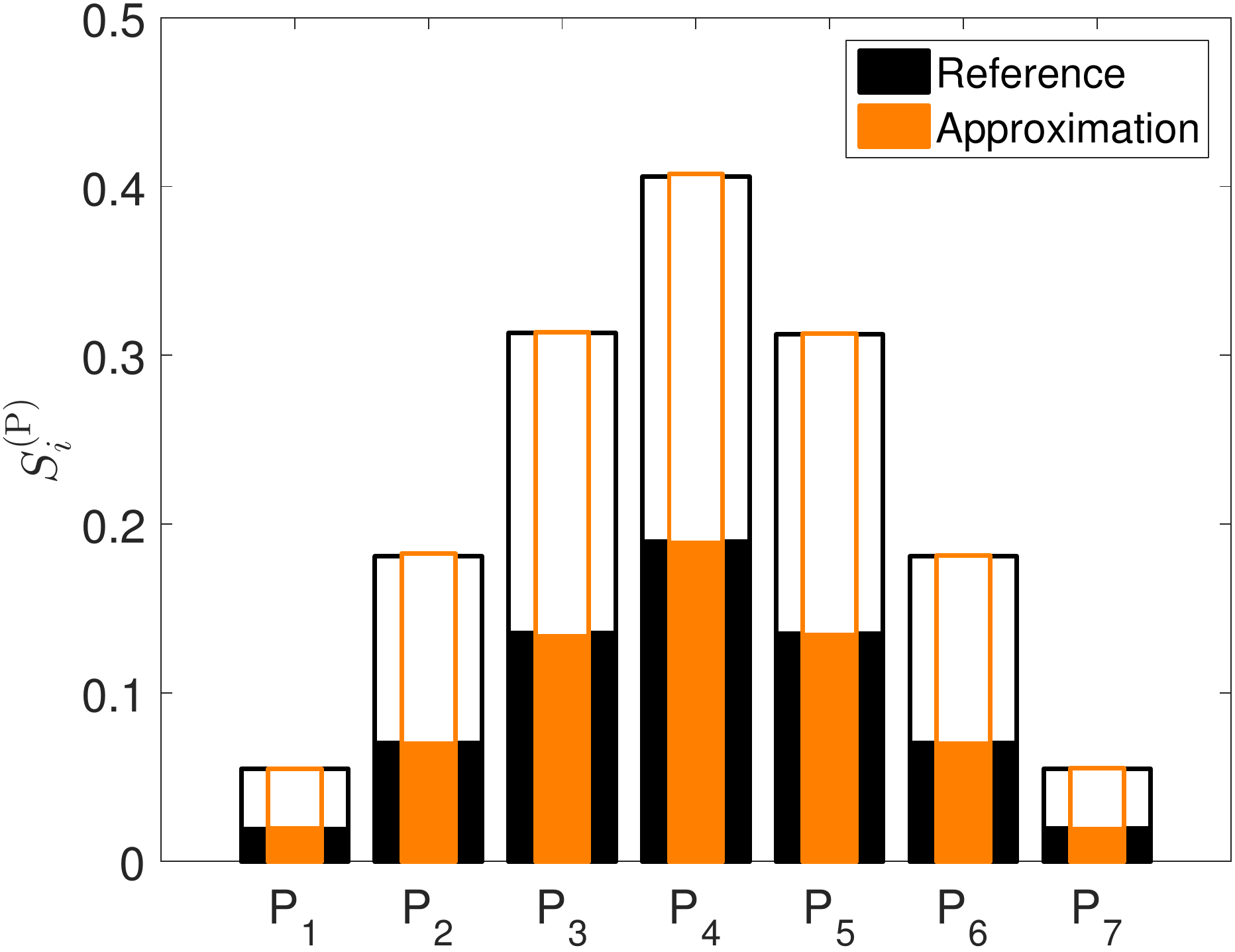}
}
\subfigure[$\nph=10$ ($\widehat{err}_{gen}=3.5\cdot 10^{-6}$)]{
\includegraphics[width=0.45\linewidth]{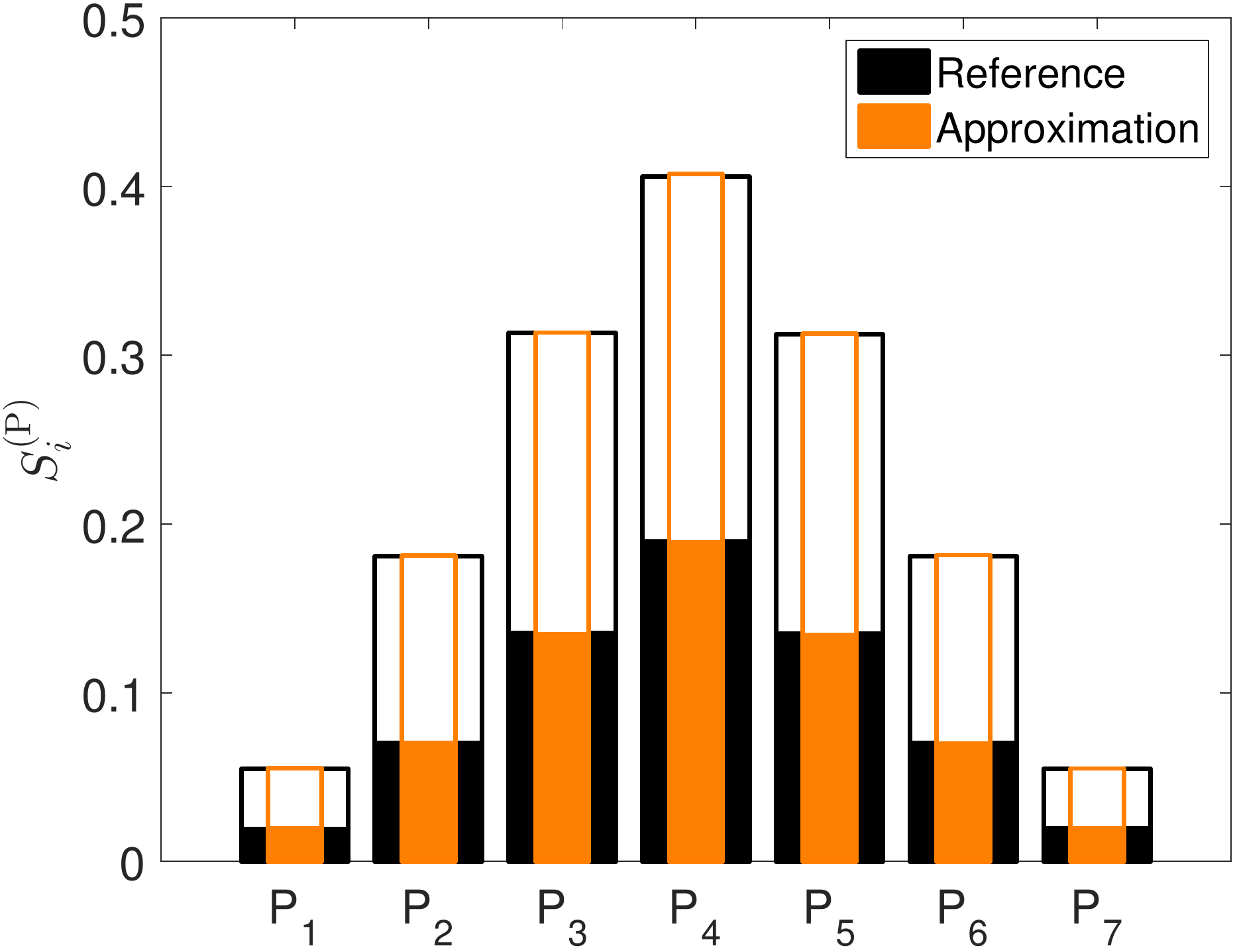}
}
\subfigure[\label{fig:struss:isobol:nph:8} $\nph=20$ ($\widehat{err}_{gen}=5.7\cdot 10^{-7}$)]{
\includegraphics[width=0.45\linewidth]{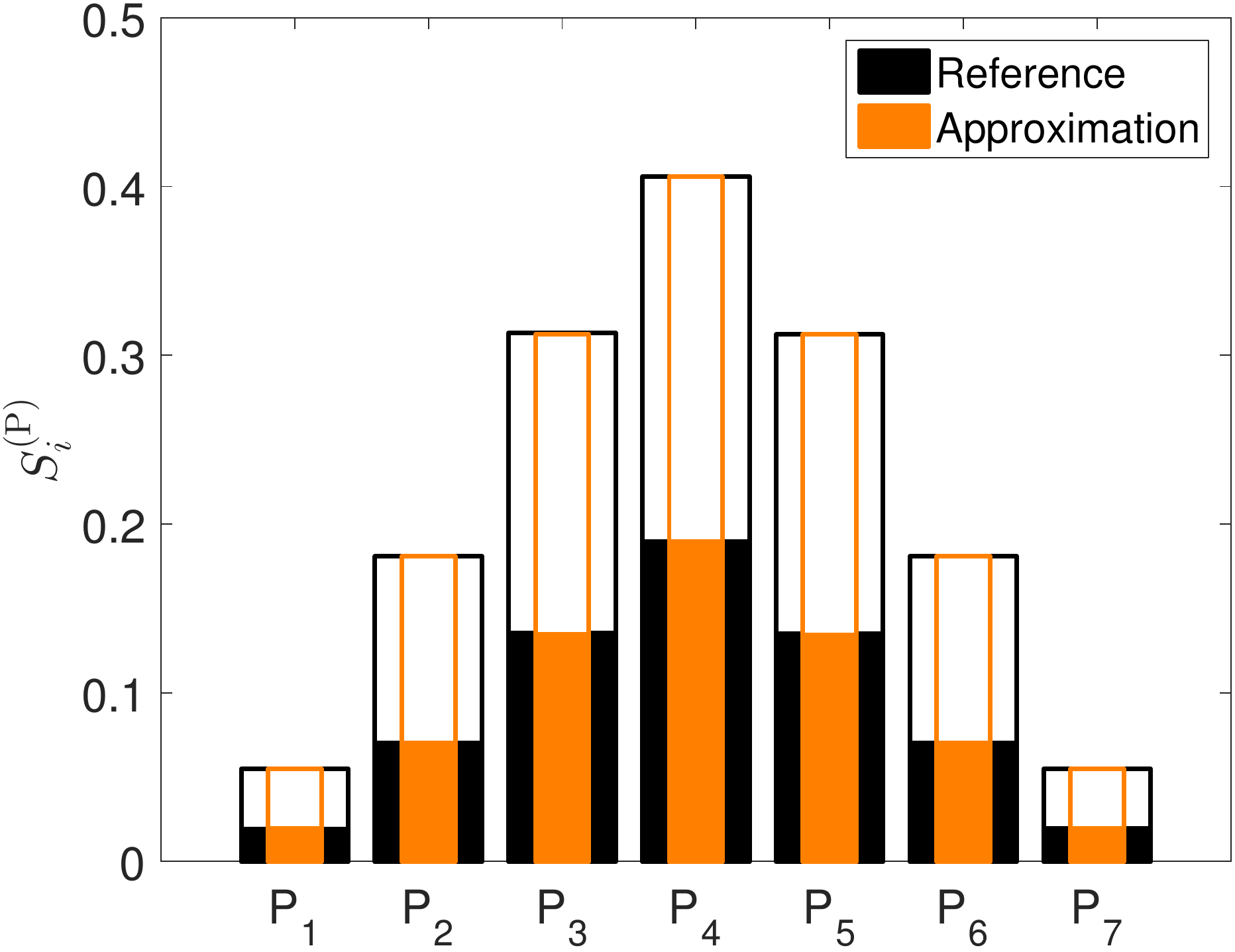}
}
\caption{Simply supported truss structure -- imprecise Sobol' indices (first order) -- augmented-PCE estimates (orange bars) versus the reference values (black bars). \label{fig:struss:isobol:nph}}
\end{figure}

The imprecise first-order Sobol' indices are visualized in Figure~\ref{fig:struss:isobol:nph} as a function of $\nph$. The filled and hollow bars represent $\underline{S}_i$ and $\overline{S}_i$, respectively. The estimated values (plotted with orange bars) are compared to the reference solution (black bars). As $\nph$ increases, the extreme values of the Sobol' indices converge to the true values without adding additional FEM runs. In this example, $\nph=5$ is sufficient to obtain reliable results. This impression is confirmed by the relative generalization error which is provided in the caption of the figures. In the case of $\nph=3$ and the corresponding relative generalization error of $\widehat{err}_{gen}=8.4\cdot 10^{-4}$, the imprecise PCE-based analysis provides usable estimates of the imprecise Sobol' indices. In total, the full set of Sobol' indices together with their ranges of epistemic uncertainty is obtained for a total cost of 100 finite element analyses of the truss structure, meaning several orders of magnitude less than any Monte Carlo-based double loop approach. 

\begin{figure}[!ht]
\centering
\includegraphics[width=0.45\linewidth]{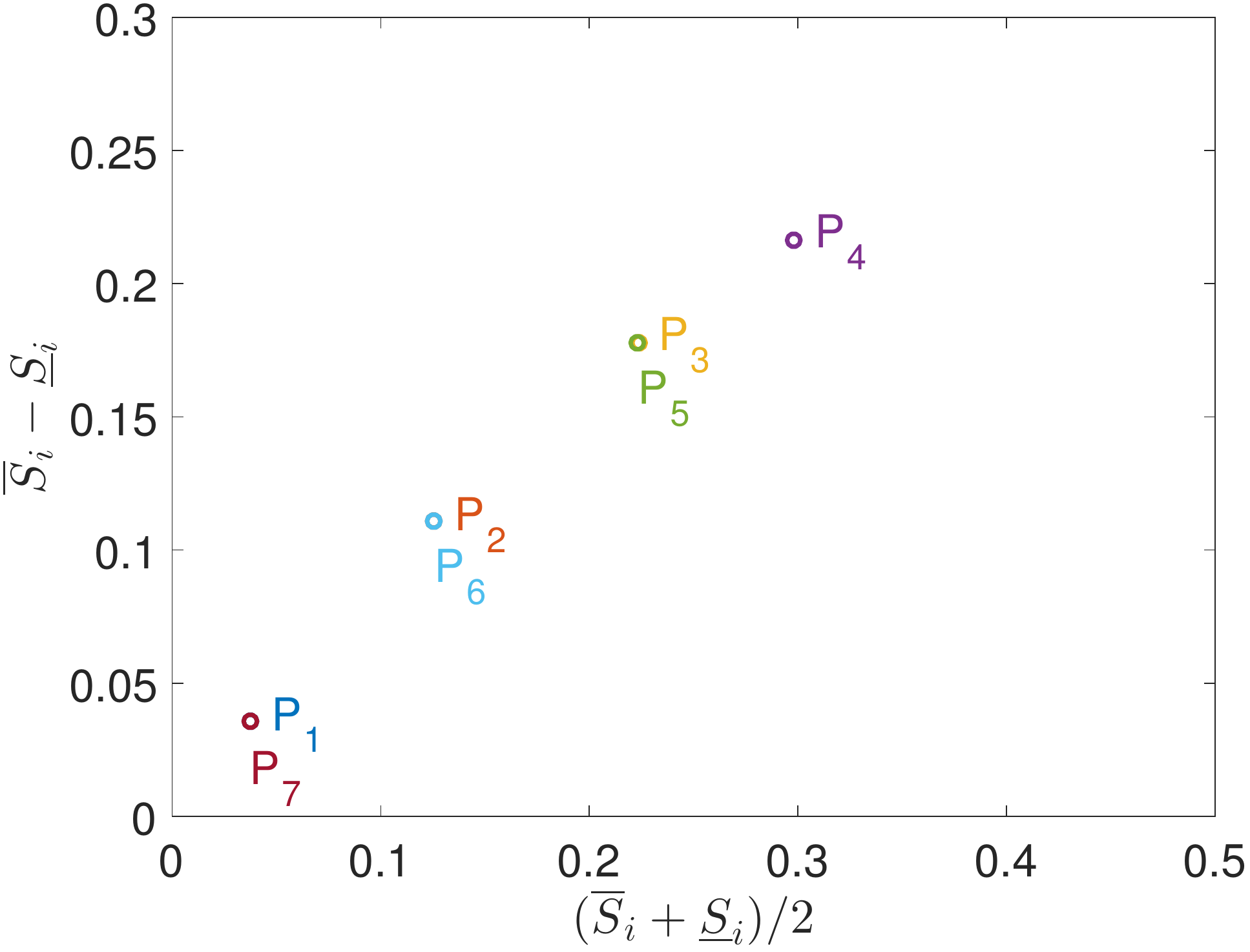}
\caption{\label{fig:truss:soso} Simply supported truss structure -- impact of aleatory vs. epistemic uncertainty plot}
\end{figure}

Similarly as for the SDOF oscillator example, Figure~\ref{fig:truss:soso} shows the impact of aleatory vs. epistemic uncertainty for the truss structure. This plot illustrates nicely the symmetry of the imprecise Sobol' indices with respect to the central load $P_4$, \ie the points representing $P_3$ and $P_5$, $P_2$ and $P_6$, $P_1$ and $P_7$ are identical. Moreover, the amount of imprecision is proportional to the absolute impact for all seven loads in this example.

\section{Conclusions} \label{sec:conc}

Sensitivity analysis is a popular tool to analyse the input-output relationship of computational models. In this paper, imprecise Sobol' indices are discussed as a measure of global sensitivity in the context of parametric probability-boxes for modelling the model input parameters. The clear separation of aleatory and epistemic uncertainty in probability-boxes allows for an intuitive interpretation of the resulting interval-valued Sobol' indices. Therein, the absolute value represents the importance of each parameter with respect to the aleatory uncertainty whereas the range of the interval represents the epistemic uncertainty. 

In order to speed up the computations of imprecise Sobol' indices, an augmented polynomial chaos expansion (PCE) model is introduced. The imprecise Sobol' indices are obtained by simply post-processing the augmented PCE model. In particular, the extreme values of the Sobol' indices are computed by optimizing on the coefficients of the augmented PCE model without rerunning the computational model. Hence, the imprecise Sobol' indices can be evaluated efficiently.  Furthermore, the introduction of phantom points counteracts the increased complexity of the augmented PCE model due to the augmented input vector.

The proposed algorithm allows for analysing realistic engineering problems as shown with the single-degree-of-freedom oscillator and the truss structure examples. A small experimental design is sufficient to estimate the imprecise PCE-based Sobol' indices accurately. Moreover, the ``phantom points trick'' allows for a further increase in meta-model accuracy and at the same time a more efficient estimation of the Sobol' indices. 

\section*{References}


\bibliographystyle{chicago}
\bibliography{biblioSOBOL}

\end{document}